\documentclass[12pt]{article}
\usepackage{amsfonts}
\usepackage{graphicx}
\usepackage{psfrag,epsfig}
\usepackage{amsmath,amssymb,amsthm,bm}
\usepackage{setspace,url} 

\usepackage{titlesec} 
\titlelabel{\thetitle.\quad} 

\usepackage{verbatim}
\usepackage{tikz}
\usepackage[round]{natbib}
\usepackage{float}
\usepackage{multicol}
\usepackage{multirow}

\usepackage{empheq}
\usepackage{enumitem}
\usepackage{adjustbox}

\RequirePackage[colorlinks,citecolor=blue,urlcolor=blue]{hyperref}

\usepackage[sc]{mathpazo}

\numberwithin{equation}{section}
\theoremstyle{plain}

\newtheorem{prop}{Proposition}

\begin{document}
\onehalfspace

\title{\textbf{Forecasting unemployment using Internet search data via PRISM}}
\date{\today}

\author{ Dingdong Yi $^{\ast }$, Shaoyang Ning $^{\ast }$, Chia-Jung Chang $^{\ast }$, S. C. Kou \thanks{
Dingdong Yi is a Quantitative Researcher at Citadal Americas LLC; email: \url{yidingdong@gmail.com}. Shaoyang Ning is Assistant Professor of Statistics, Williams College; email: \url{sn9@williams.edu}.
Chia-Jung Chang is Associate Professor, Department of Statistics and Applied Probability, National University of Singapore, Singapore; email: \url{stacc@nus.edu.sg}.
S. C. Kou is Professor, Department of
Statistics, Harvard University, Cambridge, MA 02138; email:
\url{kou@stat.harvard.edu}. }}
\date{January 2021}
\maketitle

\begin{abstract}
Big data generated from the Internet offer great potential for predictive analysis. Here we focus on using online users' Internet search data to forecast unemployment initial claims weeks into the future, which provides timely insights into the direction of the economy. To this end, we present a novel method PRISM (Penalized Regression with Inferred Seasonality Module), which uses publicly available online search data from Google. PRISM is a semi-parametric method, motivated by a general state-space formulation,  
and employs nonparametric seasonal decomposition and penalized regression. 
For forecasting unemployment initial claims, PRISM outperforms all previously available methods, including forecasting during the 2008-2009 financial crisis period and near-future forecasting during the COVID-19 pandemic period, when unemployment initial claims both rose rapidly. The timely and accurate unemployment forecasts by PRISM could aid government agencies and financial institutions to assess the economic trend 
and make well-informed decisions, especially in the face of economic turbulence.

\end{abstract}

\baselineskip22pt
\section{Introduction}

Driven by the growth and wide availability of Internet and online platforms, big data are generated with an unprecedented speed nowadays. They offer the potential to inform and transform decision making in industry, business, social policy and public health \citep{manyika2011big, mcafee2012big, chen2012business, khoury2014big, kim2014big, murdoch2013inevitable}. 
Big data can be used for developing predictive models for systems that would have been challenging to predict with traditional small-sample-based approaches \citep{einav2014data,siegel2016predictive}. 
For instance, numerous studies have demonstrated the potential of using Internet search data in tracking influenza outbreaks \citep{ginsberg2009detecting, yang2015accurate, ning2019accurate, yang2020use}, dengue fever outbreaks \citep{yang2017advances}, financial market returns \citep{preis2013quantifying,risteski2014can}, consumer behaviors \citep{goel2010predicting}, unemployment \citep{ettredge2005using, choi2012predicting,li2016nowcasting} and housing prices \citep{wu2015future}.

We consider here using Internet users' Google search to forecast US unemployment initial claims weeks into the future.
Unemployment initial claims measure the number of jobless claims filed by individuals seeking to receive state jobless benefits. 
It is closely watched by the government and the financial market, as it provides timely insights into the direction of the economy. 
A sustained increase of initial claims would indicate rising unemployment and a challenging economy, whereas a steady decrease of initial claims would signal recovery of labor market. During the great financial crises of 2008 and the COVID-19 pandemic, these unemployment data have been a key focus for government agencies when making fiscal and monetary policy decisions under unprecedented pressure. 

Weekly unemployment initial claim is the (unadjusted) total number of actual initial claims filed under the Federal-State Unemployment Insurance Program in each week ending on Saturday. 
The Employment and Training Administration (ETA) of the US Department of Labor collects the weekly unemployment insurance claims reported by each state's unemployment insurance program office, and releases the data to the public at 8:30 A.M. (eastern time) on the following Thursday. Thus, the weekly unemployment initial claim data are reported with a one-week delay: the number reported on Thursday of a given week is actually the unemployment initial claim number of the preceding week. 
For accessing the general economic trend, it is, therefore, highly desirable for government agencies and financial institutions to predict the unemployment situation of the current week, which is known as nowcasting \citep{giannone2008nowcasting}, as well as weeks into the future. In this article, we use the general phrase \emph{forecasting} to cover both nowcasting (the current week) and predicting into future weeks. 

In contrast to the one-week delayed unemployment reports by the Department of Labor, Internet users' online search of unemployment-related query terms provides highly informative and \emph{real-time} information for the current unemployment situation. For instance, a surge of Internet search of ``unemployment office'',  ``unemployment benefits'', ``unemployment extension'', etc.\ in a given week could indicate an increase of unemployment of \emph{that} week, as presumably more people unemployed are searching for information of getting unemployment aid. Internet search data, offering a real-time ``peek'' of the current week, thus, augments the delayed official time-series unemployment data. 

\begin{figure*}[!htbp]
\centering
\includegraphics[width=\textwidth]{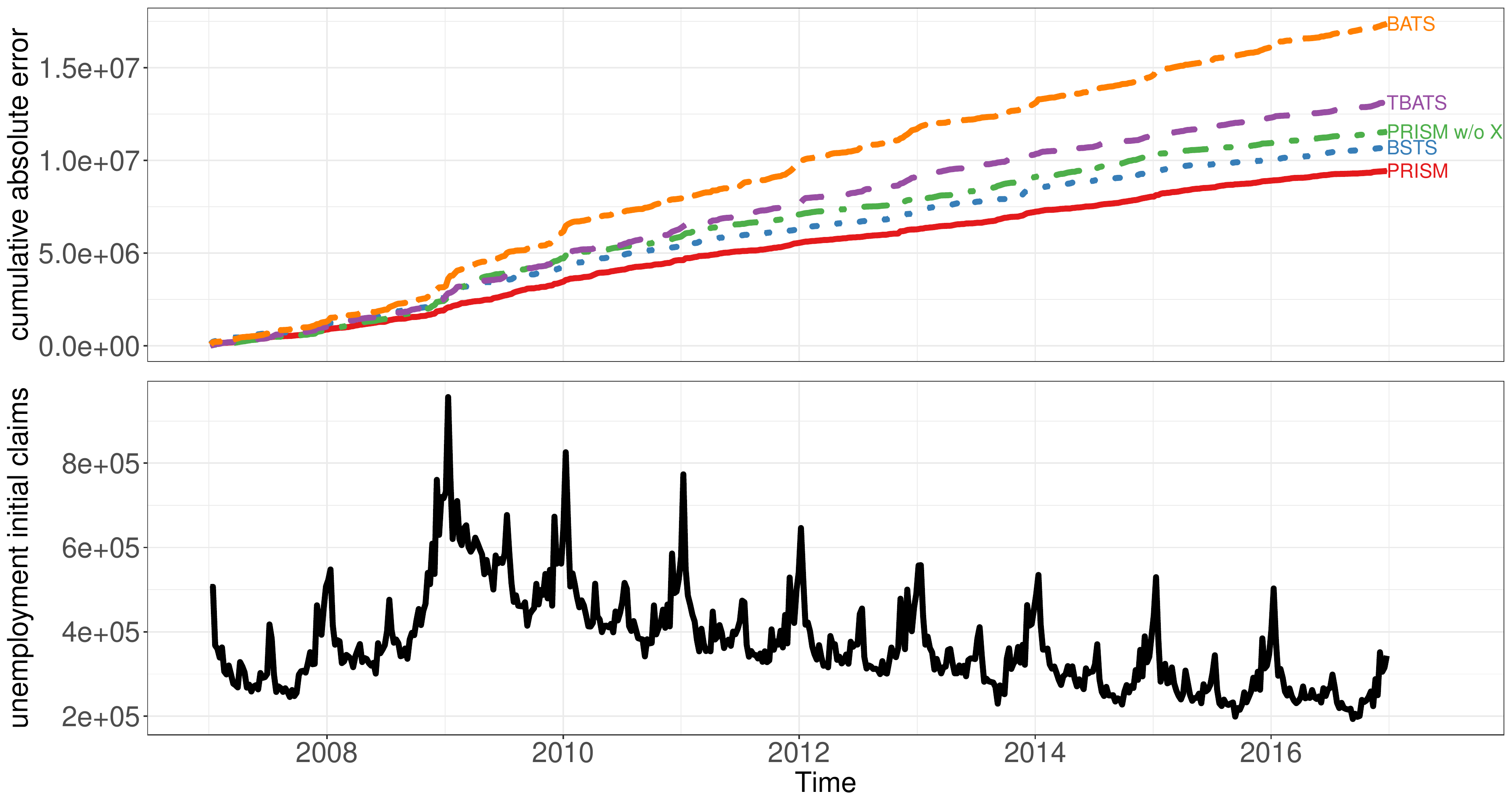}
\caption{(Top) The cumulative absolute error of nowcasting of different methods. (Bottom) The unemployment initial claims for the same period of $2007-2016$. }
\label{cum_abs_err}
\end{figure*}

There are several challenges in developing an effective method to forecast weekly unemployment initial claims with Internet search data. 
First, the volatile seasonality pattern accounts for most of the variation of the target time series. Figure \ref{cum_abs_err} (bottom) plots the weekly unemployment initial claims from 2007 to 2016; the seasonal spikes are particularly noteworthy. A prediction method should address and utilize the strong seasonality in order to achieve good prediction performance. 
Second, the method needs to effectively incorporate the most up-to-date Internet search data into the modeling of target time series.
Third, as people's search pattern and the search engine both evolve over time, the method should be able to accommodate this dynamic change.

Most time series models rely on state-space models to deal with seasonality, where the latent components capture the trend and seasonality \citep{aoki1987state, harvey1989forecasting}.
Among the time series models, structural time series models and innovation state-space models are two main frameworks \citep{harvey1993forecasting, durbin2012time, hyndman2008forecasting}, both of which have various extensions of seasonal pattern modeling and can incorporate exogenous signals as regression components (see Supplementary Material A2 for more discussion). For nowcasting time series with seasonal pattern, \cite{scott2013bayesian, scott2014predicting} developed a Bayesian method based on the structural time series model, using a spike-and-slab prior for variable selection, and applied it to nowcast unemployment initial claims with Google search data by treating the search data as regressors.
Alternative to this regression formulation, \cite{banbura2013now} proposed a nowcasting method using a factor model, in which target time series and related exogenous time series are driven by common factors. 

Here we introduce a novel prediction method PRISM, which stands for Penalized Regression with Inferred Seasonality Module, for forecasting times series with seasonality, and use it to forecast unemployment initial claims. 
Our method is semi-parametric in nature, and takes advantage of both the state-space formulation for time series forecasting and penalized regression. With the semi-parametric method PRISM, we significantly expand the range of time series models for forecasting, going beyond the traditional approaches, which are often tailored for individual parametric models. PRISM offers a robust and more accurate forecasting alternative to traditional parametric approaches 
and effectively addresses the three aforementioned challenges in forecasting time series with strong seasonality.
First, our method accommodates various nonparametric and model-based seasonal decomposition tools, and effectively incorporates the estimated seasonal components into predictive modeling. It thus can robustly handle complex seasonal patterns.
Second, different from the traditional regression formulation, our joint modeling of the target time series and the exogenous variables accommodates the potential causal relationship between them --- people do online Google search in response of becoming unemployed.
Third, PRISM uses dynamic forecasting --- training its predictive equation each week for the forecasting --- and utilizes rolling window and exponential weighting to account for the time-varying relationship between the target time series and the exogenous variables. For forecasting unemployment initial claims, PRISM delivers superior performance over all existing forecasting methods for the entire time period of $2007-2019$, and is exceptionally robust to the ups and downs of the general economic environment, including the huge volatility caused by the 2008 financial crisis.

While the forecasting target here is the unemployment initial claims, we want to highlight that PRISM applies to forecasting other time series with complex seasonal patterns.

\section{Data and Method}
\subsection{Initial Claims Data and Internet Search Data from Google}
\label{sec1.1}
The weekly (non-seasonally adjusted) initial claims are our target time series. The initial claims for the preceding week are released every Thursday. The time series of the initial claims from 1967 to present are available at \url{https://fred.stlouisfed.org/series/ICNSA}. 

The real-time Internet search data we used were obtained from Google Trends (\url{www.google.com/trends}) with Python package \texttt{pytrends}. The Google Trends website, which is publicly accessible, provides weekly (relative) search volume of search query terms specified by a user. Specifically, for a user-specified query term, Google Trends provides integer-valued weekly times series (after 2004); each number in the time series, ranging from 0 to 100, represents the search volume of that search query term in a given week divided by the total online search volume of that week; and the number is normalized to integer values from 0 to 100, where 100 corresponds to the maximum weekly search within the time period (specified by the user). Figure A1 in the Supplementary Material illustrates the Google Trend time series of several search query terms in a 5-year span. 

The search query terms in our study were also identified from the Google Trends tool. One feature of Google Trends is that, in addition to the time series of a specific term (or a general topic), it also returns the top query terms that are most highly correlated with the specific term. In our study, we used a list of top 25 Google search terms that are most highly correlated with the term ``unemployment''.  Table A1 of the Supplementary Material lists these 25 terms, which were generated by Google Trends on January 11, 2018; 
they included 12 general unemployment related query terms, such as ``unemployment office'', ``unemployment benefits'' and ``unemployment extension'', as well as 13 query terms that were combinations of state names and ``unemployment'', such as ``California unemployment'' and ``unemployment Florida''. 

\subsection{Overview of PRISM} 

PRISM employs a two-stage estimation procedure for forecasting time series $y_t$ using its lagged values and the available exogenous variables $\bm{x}_t$. The derivation and rationale of each step will be described subsequently.

\begin{itemize}[leftmargin=*]
 \item[] \textbf{Input: } Target time series $\{y_{1:(t-1)}\}$ and exogenous time series $\{\bm{x}_{t_0:t}\}$. In the forecasting of unemployment initial claims, $\{y_{1:(t-1)}\}$ is the official weekly unemployment initial claim data reported with one-week delay, and $\{\bm{x}_{t_0:t}\}$ is the multivariate Google Trends data starting from 2004.
 \item[] \textbf{Stage 1 of PRISM: Seasonal decomposition.} Decompose $\{y_t\}$, the univariate time series of interest, into the seasonal component $\{\gamma_t\}$, and the seasonally adjusted component $z_t = y_t - \gamma_t$. In particular, with a fixed rolling window length $M$, Stage 1 nonparametrically decomposes 
$\{y_{(t-M):(t-1)}\}$ into estimated seasonal component $\left\{\hat{\gamma}_{i,t}\right\}_{i=(t-M),\ldots, (t-1)}$ and estimated seasonally adjusted component $\left\{\hat{z}_{i,t}\right\}_{i=(t-M),\ldots, (t-1)}$ using data available at time $t$. 
  
\item[] \textbf{Stage 2 of PRISM: Penalized regression.} Forecast target time series using: 
\begin{equation}
\hat{y}_{t+l}=\mu_{y}^{(l)}+\sum_{j=1}^{K}\alpha_{j}^{(l)} \hat{z}_{t-j,t}+\sum_{j=1}^{K}\delta_{j}^{(l)}\hat{\gamma}_{t-j,t}+\sum_{i=1}^{p}\beta_{i}^{(l)}x_{i,t}
\label{lineareqn}
\end{equation}
where the coefficients are estimated by a rolling-window $L_1$ penalized linear regression using historical data for each forecasting horizon: $l=0$ corresponds to nowcast; $l\geq 1$ corresponds to forecasting future weeks. Note that for notational ease, we will suppress ``$(l)$'' in the superscripts of the coefficients in the subsequent discussion.
\end{itemize}

\subsection{Derivation of PRISM}

PRISM is motivated by a general state-space formulation for univariate time series with seasonal pattern. 
We postulate that the seasonal and seasonally adjusted component $\{\gamma_t\}$ and $\{z_t\}$ each evolve according to a linear state-space model with state vectors $\{\bm{s}_t\}$ and $\{\bm{h}_t\}$ respectively:
\begin{subequations}
\begin{equation}
y_t = z_t + \gamma_t
\end{equation}
\begin{minipage}{.5\textwidth}
\begin{empheq}[left=\empheqlbrace]{align}
  z_t &= \bm{w}'\bm{h}_t + \epsilon_t \\
  \bm{h}_t &= \bm{F}\bm{h}_{t-1} + \bm{\eta}_t
\end{empheq}
\end{minipage}
\begin{minipage}{.5\textwidth}
\begin{empheq}[left=\empheqlbrace]{align}
  \gamma_t &= \bm{v}'\bm{s}_t + \zeta_t \\
  \bm{s}_t &= \bm{P}\bm{s}_{t-1} + \bm{\omega}_t
\end{empheq}
\end{minipage}%
\vspace{\belowdisplayskip}
\label{uni}
\end{subequations}
where $(\epsilon_t, \zeta_t, \bm{\eta}_t', \bm{\omega}_t')' \overset{iid}{\sim} \mathcal{N}(\bm{0}, \bm{H})$, and $\bm{\theta} = (\bm{w}, \bm{F}, \bm{v}, \bm{P}, \bm{H})$ are the parameters. 

Our state-space formulation contains a variety of widely used time series models, including structural time series models \citep{harvey1989forecasting} and additive innovation state-space models \citep{aoki1987state,ord1997estimation,hyndman2008forecasting}. Under the general formulation (\ref{uni}), a specific parametric model can be obtained by specifying the state vectors $\{\bm{h}_t\}$ and $\{\bm{s}_t\}$ along with the dependence structure $\bm{H}$. We highlight a few special cases of model (\ref{uni}) in Supplementary Material.

PRISM also models the contemporaneous information from exogenous variables. Let $\bm{x}_t = (x_{1, t}, x_{2, t}, \ldots, x_{p, t})'$ be the vector of the exogenous variables at time $t$. We postulate a state-space model for $\bm{x}_t$ on top of $y_t$, instead of adding them as regressors as in traditional models. In particular, at each time $t$, we assume a multivariate normal distribution for $\bm{x}_t$ conditional on the level of unemployment initial claims $y_t$,
\begin{equation}
\bm{x}_t|y_t \sim \mathcal{N}_p(\bm{\mu}_x+y_t\bm{\beta},\bm{Q})
\label{x_multi}
\end{equation}
where $\bm{\beta}=(\beta_1,\ldots,\beta_p)'$, $\bm{\mu}_x=(\mu_{x_1},\ldots, \mu_{x_p})'$, and $\bm{Q}$ is the covariance matrix. $\bm{x}_t$ is assumed to be independent of $\{y_l , \bm{x}_l:l< t\}$ conditional on $y_t$. For $\{y_t\}$ following the general state-space model (\ref{uni}), our joint model for $y_t$ and $\bm{x}_t$ can be diagrammed as:
\begin{equation*}
\begin{array}{ccccccc}
\cdots & \rightarrow & \left(\bm{s}_t,\bm{h}_t\right) & \rightarrow & \left(\bm{s}_{t+1},\bm{h}_{t+1}\right) & \rightarrow & \cdots\\
 &  & \downarrow &  & \downarrow\\
 &  & y_{t} &  & y_{t+1}\\
 &  & \downarrow &  & \downarrow\\
 &  & \bm{x}_{t} &  & \bm{x}_{t+1}
\end{array}
\label{dg2}
\end{equation*}

To forecast $y_{t+l}$ under above model assumptions at time $t$, we consider the predictive distribution of $y_{t+l}$ by conditioning on the historical data $\{y_{1:(t-1)}\}$ and contemporaneous exogenous time series $\{\bm{x}_{t_0:t}\}$ as well as the latent seasonal component $\{\gamma_{1:(t-1)}\}$. 
$z_{1:(t-1)}$ is known given $y_{1:(t-1)}$ and $\gamma_{1:(t-1)}$. We can derive a universal representation of the predictive distribution $p (y_{t+l}\mid z_{1:(t-1)}, \gamma_{1:(t-1)}, \bm{x}_{t_0:t})$, which is normal with mean linear in $z_{1:(t-1)}$, $\gamma_{1:(t-1)}$ and $\bm{x}_t$ as in \eqref{lineareqn} (see Supplementary Material  for the proof). This observation leads to our two-stage semi-parametric estimation procedure PRISM for nowcasting $y_t$ and forecasting $y_{t+l}$ ($l\geq 1$) using all available information at time $t$.

\subsection{Stage 1 of PRISM: seasonal decomposition}

PRISM estimates the unobserved seasonal components $\gamma_{1:(t-1)}$ in the first stage. For this purpose, various seasonal decomposition methods can be used here, including nonparametric methods such as the classical additive seasonal decomposition \citep{kendall1976advanced} and parametric methods based on innovation state-space models. We use the method of Seasonal and Trend decomposition using Loess (STL) \citep{cleveland1990stl} as the default choice.
The STL method is nonparametric. It is widely used and robust for decomposing time series with few assumptions owing to its nonparametric nature. 

At every time $t$ for forecasting, we apply the seasonal decomposition method (such as the default STL) to historical initial claims observations $y_{(t-M):(t-1)}$ with $M$ being a large number. For each rolling window from $t-M$ to $t-1$, the univariate time series $y_{(t-M):(t-1)}$ is decomposed into three components: seasonal, trend and the remainder. Denote $\hat{\gamma}_{i,t}$ and $\hat{z}_{i,t}$ as the estimates of $\gamma_i$ and $z_i$ using data available at time $t$. Then, at each $t$ the seasonal decomposition generates estimated seasonal component time series $\left\{\hat{\gamma}_{i,t}\right\}_{i=(t-M),\ldots, (t-1)}$ and seasonally adjusted time series $\left\{\hat{z}_{i,t}\right\}_{i=(t-M),\ldots, (t-1)}$; the latter is the sum of trend component and remainder component. In our forecasting of unemployment initial claims, we took $M=700$. We describe the basic procedure of the default STL in the Supplementary Material.

\subsection{Stage 2 of PRISM: penalized linear regression}
For each fixed forecasting horizon $l$ ($\geq 0$), we estimate $y_{t+l}$ by the linear predictive equation:
\begin{equation}
\hat{y}_{t+l}=\mu_y+\sum_{j=1}^{K}\alpha_{j} \hat{z}_{t-j,t}+\sum_{j=1}^{K}\delta_j\hat{\gamma}_{t-j,t}+\sum_{i=1}^{p}\beta_{i}x_{i,t},
\label{sargo}
\end{equation}
where for notational ease we have suppressed $l$ in the coefficients and used the generic notations $\mu_y$, $\alpha_{j}$, $\delta_{j}$ with the understanding that there is a separate set of $\{\mu_y, \bm{\alpha}=(\alpha_1,\ldots,\alpha_K), \bm{\delta}=(\delta_1,\ldots,\delta_K), \bm{\beta}=(\beta_1,\ldots,\beta_p)\}$ for each $l$. 
At each time $t$ and for each forecasting horizon $l$, the regression coefficients $\mu_y$, $\bm{\alpha}$, $\bm{\delta}$ and $\bm{\beta}$ are obtained by minimizing
\begin{align}
\frac{1}{N}\sum_{\tau=t-l-N}^{t-l-1} & w^{t-\tau}\Big( y_{\tau+l}-\mu_y-\sum_{j=1}^{K}\alpha_{j} \hat{z}_{\tau-j,\tau}-\sum_{j=1}^{K}\delta_j\hat{\gamma}_{\tau-j,\tau} \nonumber \\
& -\sum_{i=1}^{p}\beta_{i}x_{i,\tau}\Big) ^2
 +\lambda_1\left(\|\bm{\alpha}\|_{1}+\|\bm{\delta}\|_{1}\right)+\lambda_2\|\bm{\beta}\|_{1},
\label{sargo_l1}
\end{align}
where $N$ is the length of a rolling window, $w$ is a discount factor, and $\lambda_1$ and $\lambda_2$ are nonnegative regularization parameters.

\subsection{Features of PRISM}
There are several distinct features of our estimation procedure. First, a rolling window of length $N$ is employed. This is to address the fact that the parameters in the predictive Eq.\ \eqref{sargo} can vary with time $t$. In our case, people's search pattern and the search engine tend to evolve over time, and it is quite likely that the same search phrases would contribute in different ways over time to the response variable. Correspondingly, the coefficients in \eqref{sargo} need to be estimated dynamically each week, and the more recent observations should be considered more relevant than the distant historical observations for inferring the predictive equations of current time. The rolling window of observations and the exponentially decreasing weights are utilized for such purpose. Our use of exponential weighting is related to the weighted least square formulation that is usually referred to as discounted weighted regression in the econometrics literature \citep{ameen1984discount, taylor2010exponentially}. 

Second, since the number of unknown coefficient in \eqref{sargo} tends to be quite large compared to the number of observations within the rolling window, we applied $L_1$ regularization in our rolling-window estimation \citep{tibshirani1996regression}, which gives robust and sparse estimate of the coefficients. Up to two $L_1$ penalties are applied: on $(\bm{\alpha}, \bm{\delta})$ and on $\bm{\beta}$, as they represent two sources of information: information from time series components $\left\{\hat{z}_t\right\}$ and $\left\{\hat{\gamma}_t\right\}$, and information from the exogenous variables $\{\bm{x}_t\}$.

Third, PRISM is a semi-parametric method. The predictive Eq.\ \eqref{sargo} is motivated and derived from our state-space formulation (\ref{uni}). However, the estimation is not parametric in that (i) the seasonal and seasonally adjusted components are learned non-parametrically in Stage 1, and (ii) the coefficients in \eqref{sargo} are dynamically estimated each week in Stage 2. Combined together, the two stages of PRISM give us a simple and robust estimation procedure. This approach is novel and different from the typical approaches for linear state-space models, which often estimate unknown parameters using specific parametrization and select a model based on information criteria \citep{hyndman2008forecasting}.

\subsection{Using PRISM without exogenous variables}
In the case when exogenous time series $\{\bm{x}_t\}$ are not available, PRISM estimates ${y}_{t+l}$ according to the following linear predictive equation:
\begin{equation}
\hat{y}_{t+l}=\mu_y+\sum_{j=1}^{K}\alpha_{j} \hat{z}_{t-j,t}+\sum_{j=1}^{K}\delta_j\hat{\gamma}_{t-j,t},
\label{sar2}
\end{equation}
which is a degenerated special case of the predictive Eq.\ \eqref{sargo}. 
Under the same estimation procedure as in \eqref{sargo_l1} except that $\bm{\beta}$ and $\bm{x}_t$ are dropped, predictive Eq.\ \eqref{sar2} can be used to forecast univariate time series with seasonal patterns without exogenous time series.

\subsection{Constructing point-wise predictive intervals for PRISM estimate}

The semi-parametric nature of PRISM makes it more difficult to construct predictive intervals on PRISM forecasts, as we cannot rely on parametric specifications, such as posterior distributions, for predictive interval construction. However, the fact that we are forecasting time series suggests a (non-parametric) method for us to construct predictive intervals based on the historical performances of PRISM. 

For nowcasting at time $t$, given the historical data available up to time $t-1$, we can evaluate the root mean square error of nowcasting for the last $L$ time periods as 
$$\hat{\text{se}}_t = \big( \frac{1}{L}\sum_{\tau=t-L}^{t-1}(\hat{y}_{\tau}-y_{\tau})^2 \big)^{1/2},$$
where $\hat{y}_{\tau}$ is the real time PRISM estimate for $y_\tau$ generated at time $\tau$. Under the assumption of local stationarity and normality of the residual, $\hat{\text{se}}_t$ would be an estimate for the standard error of $\hat{y}_t$. We can thus use it to construct predictive interval for the current PRISM estimate. An $1- \alpha$ point-wise predictive interval is given by $(\hat{y}_t - z_{\alpha/2} \ \hat{\text{se}}_t,\  \hat{y}_t + z_{\alpha/2} \ \hat{\text{se}}_t)$, where $z_{\alpha/2}$ is the $1- \alpha/2$ quantile of the standard normal distribution. Supplementary Material A10 shows that the empirical residuals are approximately normal, supporting our construction. The point-wise predictive intervals for forecasting into future weeks can be constructed similarly. In practice, we set $L=52$, estimating the $\hat{\text{se}}_t$ based on the forecasts of the most recent one-year window.

\subsection{Training PRISM for forecasting unemployment initial claims}

In our forecasting of unemployment initial claims, we applied a 3-year rolling window of historical data to estimate the parameters in \eqref{sargo_l1}, i.e. $N=156$ (weeks). The choice of 3-year rolling window is recommended in the literature \citep{d2017predictive} as well as supported by our empirical studies (Supplementary A7). In addition, since the Google Trends data are only available since 2004, with the 3-year rolling window we are able to test the performance of PRISM in 2007-2009, the entire span of the financial crisis, which serves as an important test of the capability of the various prediction methods. We took $K=52$ (weeks) to employ the most recent 1-year estimated seasonal and seasonally adjusted components, and $p=25$ (Google search terms) according to the list of top 25 nationwide query terms related to ``unemployment'' in Table A1. 

The weekly search volume data from Google Trends are limited up to a 5-year span per download. The subsequent normalization of the search volumes done by Google is based on the search query term and the specified date range, which implies that the absolute values of search volumes are normalized differently between different 5-year ranges. However, within the same set of 5-year data the relative scale of variables is consistent (as they are normalized by the same constant). Therefore, to avoid the variability from different normalization across different 5-year spans, and to ensure the coherence of the model, for each week, we used the same set of 5-year-span data downloaded for both the training data and the prediction.

For the choice of the discount factor, we took $w= 0.985$ as the default choice. This follows the suggestion by \cite{lindo1997optimal} that setting the discount factor between 0.95 and 0.995 works in most applications. We further conducted the experiments for $w \in [0.95, 0.995]$ (see Supplementary Material A6) and found the performance of PRISM is quite robust for $w \in [0.95, 0.995]$ while $w=0.985$ gives optimal accuracy for our in-sample nowcasting.

For the regularization parameters $\lambda_1$ and $\lambda_2$ in \eqref{sargo_l1}, we used cross-validation to minimize the mean squared predictive errors (i.e., the average of prediction errors from each validation set of data). We found empirically that the extra flexibility of having two separate $\lambda_1$ and $\lambda_2$ does not give improvement over fixing $\lambda_{1} = \lambda_{2}$. In particular, we found that for every forecasting horizon $l = 0,1,2,3$, in the cross-validation process of setting $(\lambda_1, \lambda_2)$ for separate $L_1$ penalty, over 80\% of the weeks showed that the smallest cross-validation mean squared error when restricting $\lambda_1 = \lambda_2$ is within 1 standard error of the global smallest cross-validation mean squared error. For model simplicity, we thus chose to further restrict $\lambda_1=\lambda_2$ when forecasting unemployment initial claims. 

\subsection{Accuracy metrics}
We used root-mean-squared error (RMSE) and mean absolute error (MAE) to evaluate the performance of different methods. For an estimator $\{\hat{y}_{t}\}$ and horizon $l$, the RMSE and MAE are defined, respectively, as $\text{RMSE}_l (\bm{\hat{y}}, \bm{y}) = [\frac{1}{(n_2-n_1-l+1)}\sum_{t=n_1+l}^{n_2}(\hat{y}_t - y_t)^2]^{1/2}$ and $\text{MAE}_l (\bm{\hat{y}}, \bm{y}) = \frac{1}{(n_2-n_1-l+1)}\sum_{t=n_1+l}^{n_2}|\hat{y}_t - y_t|$, where $n_1+l$ and $n_2$ are respectively the start and end of the forecasting period for each $l$.

\section{Results}
\subsection{Retrospective Forecasting for 2007-2016}
We applied PRISM to produce forecasts of weekly unemployment initial claims for the time period of 2007 to 2016 for four time horizons: real-time, one, two, and three weeks ahead of the current time. We compared the forecasts to the ground truth --- the unemployment initial claims released by the Department of Labor one-week behind real-time --- by measuring the RMSE and MAE.

For comparison, we calculated the RMSE and MAE of four alternative forecasting methods: (a) Bayesian Structural Time Series (BSTS) \citep{scott2014predicting}; (b) and (c), two forecasting methods using exponential smoothing: BATS and TBATS \citep{de2011forecasting}; and (d) the naive method, which without any modeling effort simply uses the last available weekly unemployment initial claims number (which is of the prior week) as the prediction for the current week, one, two, and three week(s) later. The naive method serves as a baseline. Both BATS and TBATS are based on innovation state-space model; BATS is an acronym for key features of the model: Box-Cox transformation, ARMA errors, Trend, and Seasonal components; TBATS extends BATS to handle complex seasonal patterns with trigonometric representations, and the initial T connotes ``trigonometric''. BSTS only produces nowcasting; it does not produce numbers for forecasting into future weeks.

As PRISM uses both historical unemployment initial claims data and Google search information, to quantify the contribution of the two resources, we also applied PRISM but without the Google search information. We denoted this method as ``PRISM w/o $\bm{x}_t$''. 

For fair comparison, in generating retrospective estimates of unemployment initial claims, we reran all methods each week using only the information available up to that week, i.e., we obtained the retrospective estimation as if we had relived the testing period of $2007-2016$. The two exponential smoothing methods (b) BATS and (c) TBATS only use historical initial claims data and do not offer the option of including exogenous variable in their forecasting, while the method (a) BSTS allows the inclusion of exogenous variables. Thus, for forecasting at each week $t$, BSTS takes the historical initial claim data and Google Trends data as input, whereas BATS and TBATS use historical initial claim data only. PRISM was applied twice: with and without Google search information. 
The results of BSTS, BATS and TBATS were produced by their respective R packages under their default settings.

Table \ref{overall} presents the performance of forecasting (including nowcasting) unemployment initial claims over the entire period of $2007-2016$ for the four forecasting horizons. The RMSE and MAE numbers reported here are relative to the naive method, i.e., the number reported in each cell is the ratio of the error of a given method to that of the naive method. The absolute error of the naive method is reported in the parentheses. BSTS does not produce numbers for forecasting into future weeks, as its R package outputs prediction of the target time series only with exogenous variables inputted.

Table \ref{overall} reveals the following. First, PRISM uniformly outperforms all the other methods for the entire period of $2007-2016$ under all forecasting horizons. Second, the real-time Google Trends data are very helpful for nowcasting, as PRISM and BSTS have better nowcasting results than the other methods that use only historical initial claim data. Third, the contribution of contemporaneous Google Trends data becomes less significant in forecasting future weeks, as evidenced by the shrinking performance gap between PRISM and ``PRISM w/o $\bm{x}_t$''  from nowcasting to forecasting. Fourth, among the three methods that only use historical initial claim data, the predictive method based on PRISM without Google information outperforms the exponential smoothing methods BATS and TBATS.

Following the suggestion of a referee, we further compared the performance of PRISM to the other methods with an additional metric: Cumulative Sum of Squared forecast Error Differences (CSSED) \citep{welch2008comprehensive}, which calculates the cumulative difference in mean-squared error (MSE) between PRISM and the alternative. The CSSED at time $T$ for an alternative method $m$ is defined as $CSSED_{m, \text{PRISM}}=\sum_{t=1}^T(e_{t,m}^2-e_{t,\text{PRISM}}^2)$, where $e_{t,m}$ and $e_{t,\text{PRISM}}$ are the prediction errors at time $t$ for method $m$ and PRISM respectively. The detailed comparison results are given in Supplementary Material A13, which again shows that the advantage of PRISM over alternatives is consistent over the whole evaluation period.

\begin{table}[!htbp]
\footnotesize
\begin{center}

\begin{adjustbox}{center}
\begin{tabular}{lllllll}
\hline
 & real-time & 1 week & 2 weeks & 3 weeks \\ 
\hline
RMSE\\

\qquad PRISM & \textbf{0.493} & \textbf{0.483} & \textbf{0.461} & \textbf{0.470} \\ 
\qquad PRISM w/o $\bm{x}_t$ & 0.647 & 0.532 & 0.507 & 0.524 \\
\qquad BSTS & 0.588 & - & - & - \\ 
\qquad BATS & 1.002 & 0.897 & 0.848 & 0.832 \\ 
\qquad TBATS & 0.711 & 0.559 & 0.544 & 0.528 \\ 
\qquad naive & 1 (50551) & 1 (62227) & 1 (69747) & 1 (73527) \\ 
\\
MAE\\
\qquad PRISM & \textbf{0.539} & \textbf{0.517} & \textbf{0.476} & \textbf{0.460} \\ 
\qquad PRISM w/o $\bm{x}_t$ & 0.659 & 0.559 & 0.510 & 0.496 \\ 
\qquad  BSTS & 0.612 & - & - & - \\ 
\qquad  BATS & 0.992 & 0.898 & 0.825 & 0.781 \\ 
\qquad  TBATS & 0.750 & 0.599 & 0.570 & 0.525 \\ 
\qquad  naive & 1 (33637) & 1 (41121) & 1 (47902) & 1 (52794) \\    \hline
\normalsize
\end{tabular}
\end{adjustbox}
\caption{Performance of different methods over 
$2007-2016$ for four forecasting horizons: real-time, 1 week, 2 weeks and 3 weeks ahead. RMSE and MAE here are relative to the error of naive method; that is, the number reported is the ratio of the error of a given method to that of the naive method; the absolute RMSE and MAE of the naive method are reported in the parentheses. The boldface indicates the best performer for each forecasting horizon and each accuracy metric.}
\label{overall}
\end{center}
\end{table}

To assess the statistical significance of the improved prediction power of PRISM compared to the alternatives, we conducted Diebold-Mariano test \citep{Diebold1995}, which is a nonparametric test for comparing the prediction accuracy between two time-series forecasting methods. 
Table \ref{dmtest} reports the p-values of the Diebold-Mariano test (the null hypothesis being that PRISM and the alternative method in comparison have the same prediction accuracy in RMSE). With all the p-values smaller than 2.1\%, Table \ref{dmtest} shows that the improved prediction accuracy of PRISM over BSTS, BATS, and TBATS is statistically significant in all of the forecasting horizons evaluated. Further comparison with two additional methods is presented in Supplementary Material A12, where PRISM continues to show significant advantage in prediction accuracy over seasonal AR model and the method of \cite{d2017predictive}. 

\begin{table}[!h]
\footnotesize
\begin{center}
\begin{adjustbox}{center}
\begin{tabular}{llllll}
 \hline
 & real-time & 1 week & 2 weeks & 3 weeks \\ 
  \hline
\qquad PRISM w/o & $7.86 \times 10^{-5}$ & $2.09 \times 10^{-2}$  & $2.10 \times 10^{-2}$ & $3.95\times 10^{-3}$ \\ 
\qquad  BSTS  & $7.14 \times 10^{-3}$ &     - &     - &     - \\ 
\qquad  BATS & $9.17 \times 10^{-8}$ & $3.60 \times 10^{-8}$ & $1.05 \times 10^{-9}$ & $1.88 \times 10^{-9}$ \\ 
\qquad  TBATS & $5.20 \times 10^{-9}$ & $7.24\times 10^{-3}$  & $2.59 \times 10^{-3}$ &  $1.86 \times 10^{-2}$\\ 
   \hline
   \normalsize
\end{tabular}
\end{adjustbox}
\caption{P-values of the Diebold-Mariano test for prediction accuracy comparison between PRISM and the alternatives over 2007--2016 for four forecasting horizons: real-time, 1 week, 2 weeks and 3 weeks ahead. The null hypothesis of the test is that PRISM and the alternative method in comparison have the same prediction accuracy in RMSE.
}
\label{dmtest}
\end{center}
\end{table}

\begin{figure}[!h]
\centering
\includegraphics[width=\textwidth]{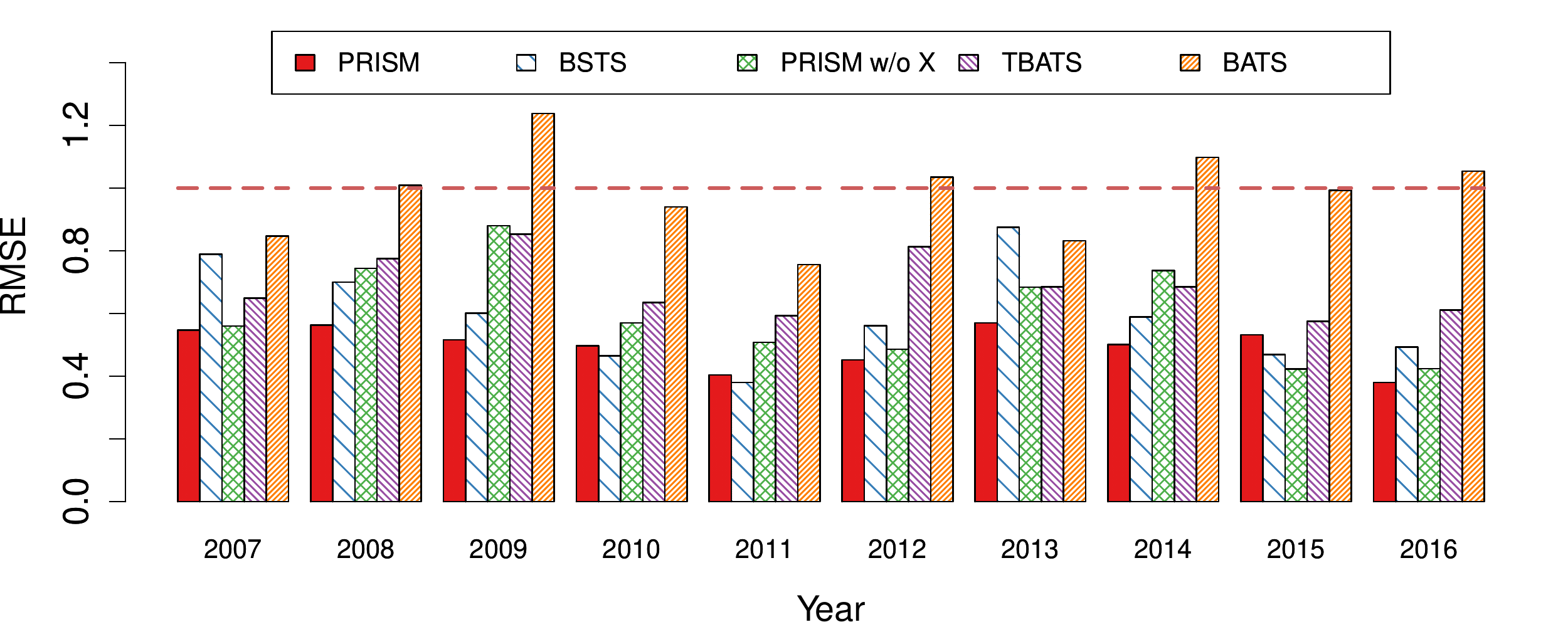}\caption{Yearly nowcasting performance of different methods from 2007 to 2016. RMSE is measured relative to the error of the naive method; a value above 1 indicates that the method performs worse than the naive method in that time period.}
\label{rmse_by_year}
\end{figure}

Figure \ref{rmse_by_year} shows the RMSE of the yearly nowcasting results of the different methods; here the RMSE is measured relative to the error of the naive method. It is seen that
PRISM gives consistent relative RMSE throughout the $2007-2016$ period. It is noteworthy that PRISM outperforms all other methods in 2008 and 2009 when the financial crisis caused significant instability in the US economy. For predictions into future weeks, PRISM also gives the leading performance for the $2007-2016$ period (detailed year-by-year plots for the forecasting performance of the different methods for each forecasting horizon are provided in Supplementary Material A14).

For a closer look of the performance of different methods, Figure \ref{cum_abs_err} (top) shows how the absolute errors of nowcasting accumulate through 2007 to 2016. The cumulative absolute error of PRISM rises at the slowest rate among all methods. As shown in Figure \ref{cum_abs_err} (bottom), the 2008 financial crisis caused significantly more unemployment initial claims. PRISM handles the financial crisis period well, as the accumulation of error is rather smooth for the financial crisis period. Other methods all accumulate loss in a considerably higher rate during the financial crisis. Furthermore, PRISM handles the strong seasonality of initial claim data well, since the accumulation of error is smooth within each year. Among all the methods considered, BATS is bumpy in handling seasonality, as the accumulation jumps when the initial claim data spikes. 

\begin{figure*}[!h]
\centering
\includegraphics[width=\textwidth]{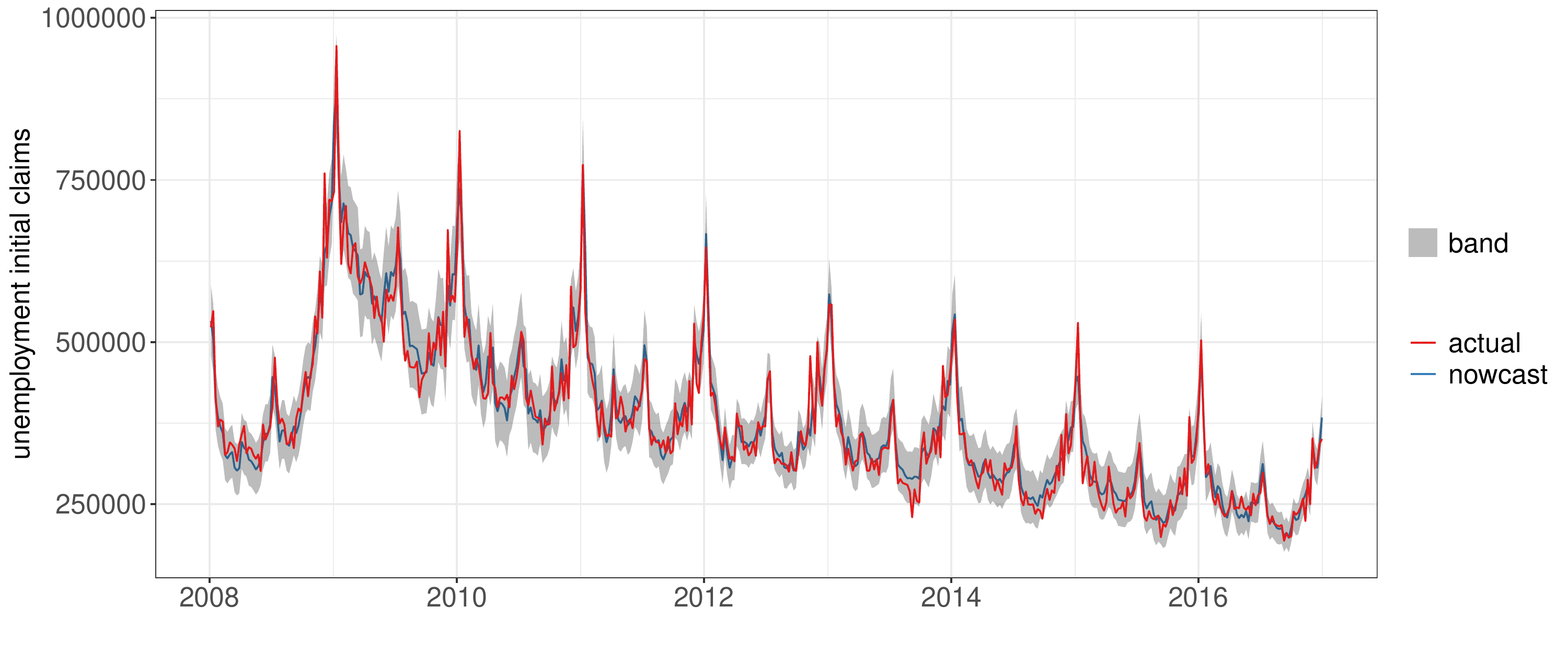}
\caption{Predictive Interval of PRISM from 2008 to 2016. The shaded area corresponds to the 95\% point-wise predictive interval of PRISM nowcasting. The blue curve is the point estimate of PRISM nowcasting. The red curve is the true unemployment initial claims. The actual coverage of the 95\% PRISM predictive interval is 96.6\% in $2008-2016$.}
\label{sargo_err_bar}
\end{figure*}

We further constructed point-wise predictive intervals for the PRISM estimates. Figure \ref{sargo_err_bar} shows the point estimates and 95\% predictive intervals by PRISM for the nowcasting during $2008-2016$ in comparison to the true unemployment initial claims officially revealed a week later (in red). 
For $2008-2016$, the actual coverage of the predictive interval is 96.6\%, which is slightly higher than the nominal 95\%. For longer forecasting horizons, the predictive intervals by PRISM also give coverage close to the nominal 95\%: for the one-week-ahead, two-week-ahead and three-week-ahead forecasts, the actual coverage levels of the PRISM predictive intervals are, respectively, 93.9\%, 95.4\% and 94.7\% (detailed plots of the PRISM predictive intervals are provided in Supplementary Material A15).

\subsection{Out-of-Sample Performance 2017-2019}
To further assess PRISM's performance, we applied PRISM to produce out-of-sample forecasts of weekly unemployment initial claims for the period of 2017-2019. Note that the PRISM methodology, including all the model specifications, was frozen at the end of 2016, so this evaluation is completely out of sample.

\begin{table}[!h]
\footnotesize
\begin{center}
\begin{adjustbox}{center}
\begin{tabular}{lllllll}
\hline
 & real-time & 1 week & 2 weeks & 3 weeks \\ 
\hline
RMSE\\
\qquad PRISM & \textbf{0.497} & \textbf{0.442} & \textbf{0.376} & \textbf{0.343} \\  
\qquad PRISM w/o $\bm{x}_t$ & 0.550 & 0.454 & 0.387 & 0.349 \\ 
\qquad   BSTS & 0.921 & - & - & - \\ 
\qquad   BATS & 1.064 & 0.982 & 0.907 & 0.832 \\ 
\qquad   TBATS & 0.699 & 0.581 & 0.512 & 0.465 \\ 
\qquad  naive & 1(27941) & 1(34936) & 1(41651) & 1(46749) \\ 

MAE\\
 \qquad PRISM &\textbf{0.550} & \textbf{0.484} & \textbf{0.411} & 0.368 \\ 
\qquad PRISM w/o $\bm{x}_t$ & 0.617 & 0.502 & 0.422 & \textbf{0.365} \\ 
 \qquad  BSTS & 0.941 & - & - & - \\ 
 \qquad  BATS & 1.155 & 0.972 & 0.907 & 0.816 \\ 
 \qquad TBATS & 0.694 & 0.587 & 0.505 & 0.435 \\ 
\qquad  naive & 1(19686) & 1(24630) & 1(29971) & 1(35016) \\ 
  \hline
\normalsize
\end{tabular}
\end{adjustbox}
\caption{Performance of different methods over $2017-2019$ for four forecasting horizons: real-time, 1 week, 2 weeks and 3 weeks ahead. RMSE and MAE here are relative to the error of naive method; that is, the number reported is the ratio of the error of a given method to that of the naive method; the absolute RMSE and MAE of the naive method are reported in the parentheses. The boldface indicates the best performer for each forecasting horizon and each accuracy metric.}
\label{overall2019}
\end{center}
\end{table}

\begin{table}[!h]
\footnotesize
\begin{center}
\begin{adjustbox}{center}
\begin{tabular}{llllll}
 \hline
 & real-time & 1 week & 2 weeks & 3 weeks \\ 
  \hline
  \qquad PRISM w/o $\bm{x}_t$  & $2.74 \times 10^{-9}$ & $4.26 \times 10^{-5}$ & $1.25 \times 10^{-5}$ &  $2.70 \times 10^{-2}$ \\ 
\qquad BSTS & $2.64 \times 10^{-5}$ & - & - & - \\ 
 \qquad BATS & $3.80 \times 10^{-7}$ & $3.09 \times 10^{-5}$ & $2.40 \times 10^{-5}$ & $4.59 \times 10^{-6}$ \\ 
\qquad  TBATS & $1.04 \times 10^{-3}$ & $2.79 \times 10^{-3}$  &  $7.46 \times 10^{-3}$  &   $2.48 \times 10^{-3}$\\
   \hline
   \normalsize
\end{tabular}
\end{adjustbox}
\caption{P-values of the Diebold-Mariano test for prediction accuracy comparison between PRISM and the alternatives over 2017--2019 for four forecasting horizons: real-time, 1 week, 2 weeks and 3 weeks ahead. The null hypothesis of the test is that PRISM and the alternative method in comparison have the same prediction accuracy in RMSE.}
\label{dmtest1719}
\end{center}
\end{table}

\begin{figure}[!htbp]
\centering
\includegraphics[width=\textwidth]{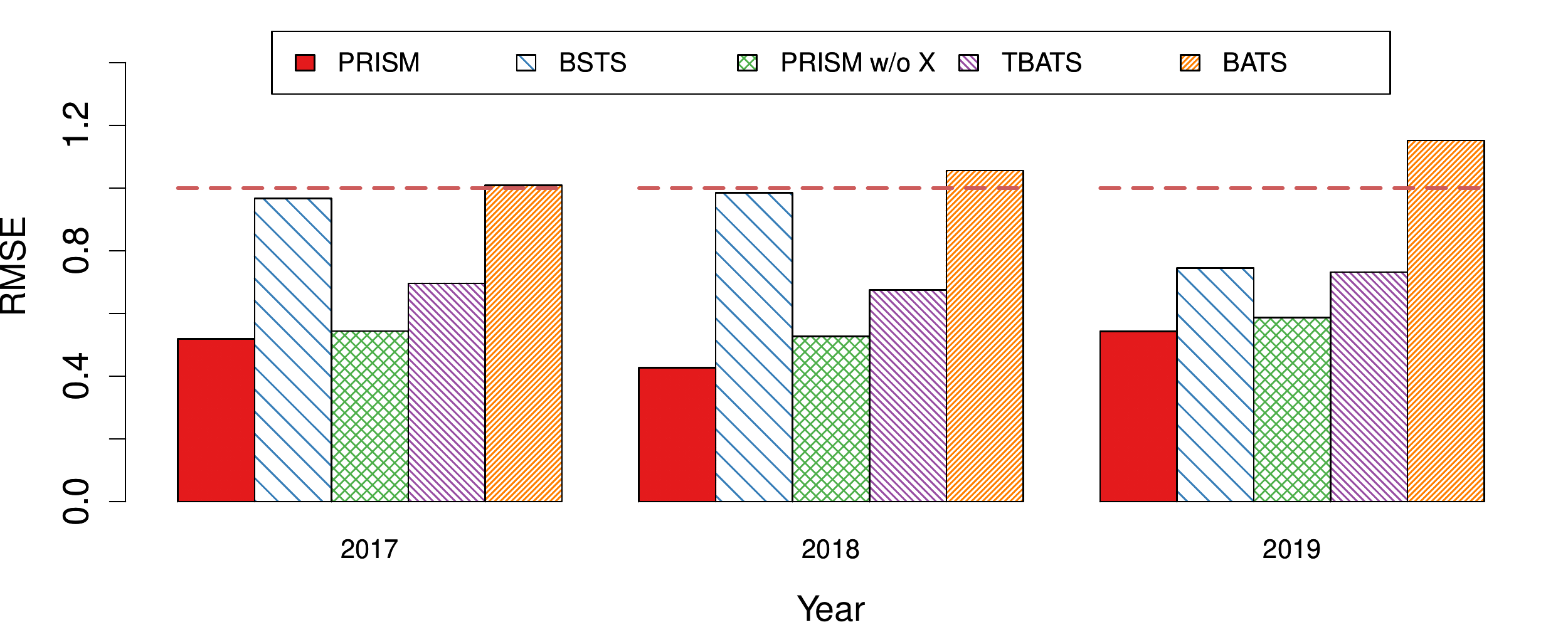}
\caption{Yearly nowcasting performance of different methods from 2017 to 2019. RMSE is measured relative to the error of the naive method; a value above 1 indicates that the method performs worse than the naive method in that time period.}
\label{rmse_by_year1719}
\end{figure}

Table \ref{overall2019} summarizes the prediction errors of PRISM (both with and without Google data) in both RMSE and MAE in comparison with other benchmark methods. PRISM again shows consistent advantage over other benchmark methods in out-of-sample predictions.  PRISM with Google data is leading across the board (except for the MAE of 3-week-ahead prediction where it virtually ties with PRISM without Google information). Notably, the relative errors compared with the naive method are quite stable over the years, similar to the results in 2007-2016. Breaking into each year (as shown in Figure \ref{rmse_by_year1719}), PRISM uniformly outperforms other methods in comparison and gives rather stable error reduction from the naive method over the years, both in- and out-of-sample. This figure together with Figure \ref{rmse_by_year} shows that PRISM reduced around 50\% error from the native method (in RMSE) year across year from 2007 to 2019. The statistical significance of PRISM's improved prediction power is also verified by the Diebold-Mariano test in Table \ref{dmtest1719}, where all the p-values are  smaller than 3\%. The consistent performance of PRISM both in the retrospective testing of 2007-2016 and the out-of-sample testing of 2017-2019 indicates the robustness and accuracy of PRISM over changes in economic environments and trends. 

\begin{figure*}[!htbp]
\centering
\includegraphics[width=\textwidth]{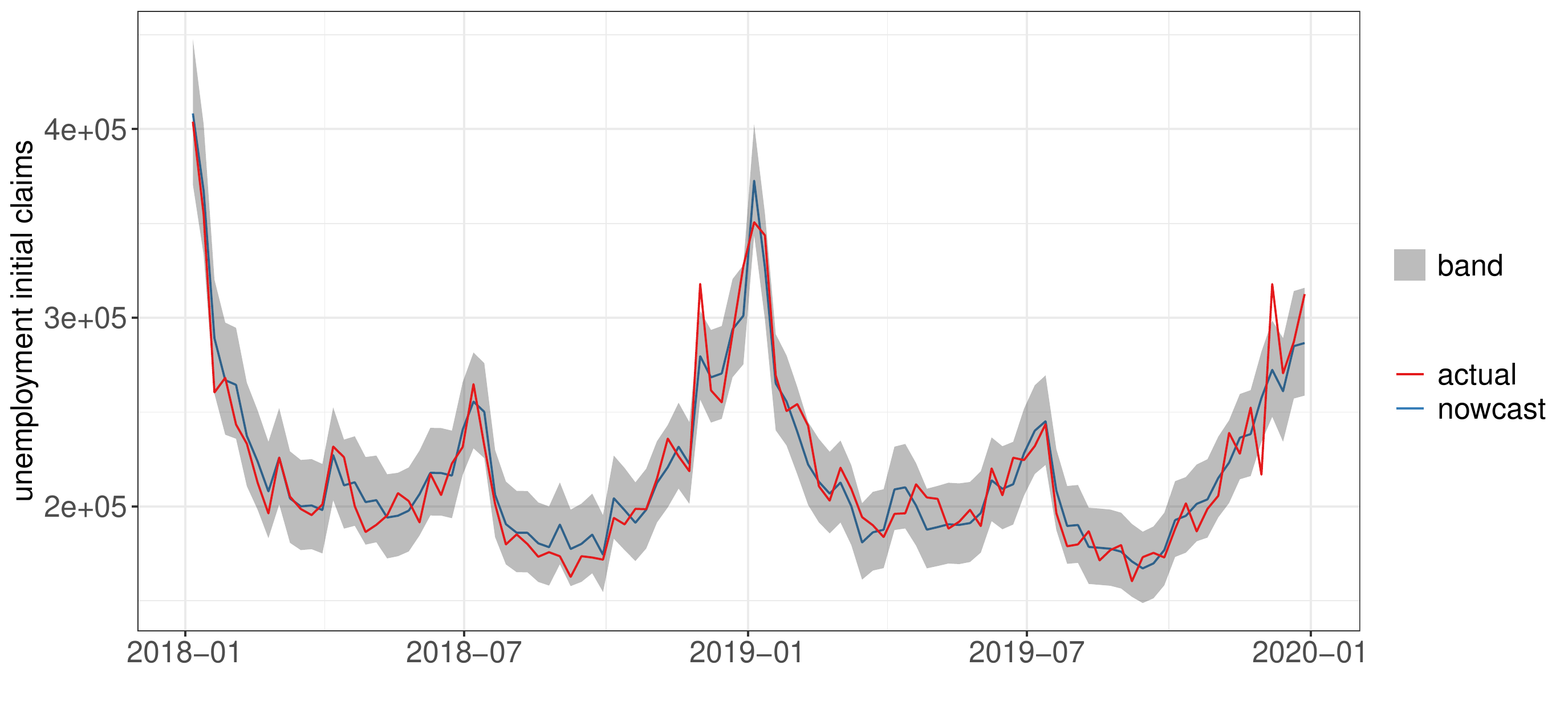}
\caption{Predictive Interval of PRISM from 2017 to 2019. The shaded area corresponds to the 95\% point-wise predictive interval of PRISM nowcasting. The blue curve is the point estimate of PRISM nowcasting. The red curve is the true unemployment initial claims. The actual coverage of the 95\% PRISM predictive interval is 97.1\% in $2017-2019$.}
\label{sargo_err_bar1719}
\end{figure*}

We also constructed point-wise predictive intervals based on the out-of-sample nowcasts in 2017-2019 (Figure \ref{sargo_err_bar1719}). Compared with the actual unemployment data released one weak later by the Department of Labor, the intervals by PRISM capture the actual numbers of unemployment initial claims in 97.1\% of the weeks in $2017-2019$, which is higher than the nominal 95\% level. This again underscores the stability of the PRISM methodology.

\subsection{Out-of-Sample Performance During COVID-19 Pandemic Period}\label{covid}

The global shutdown due to the COVID-19 pandemic has heavily impacted the US economy and job market. In particular, the numbers of unemployment claims in the US have skyrocketed to record-breaking levels with more than 40 millions people in total filing for initial claims since the start of the pandemic. This phenomenon has attracted significant attention from major news media and the general public \citep{cohen2020reversal, cohen2020million, casselman2020stubborn}. As the weekly unemployment claims remain "stubbornly high" \citep{casselman2020stubborn}, concerns for significant layoffs and severe economic downturn persist. Accurate and reliable forecasting of near-future unemployment claims would thus provide very valuable insights into the trend of the general economy during such trying times. In light of this, we further applied PRISM to the out-of-sample data of the COVID-19 pandemic period to evaluate its performance and robustness to such unusual economic shock.

\begin{figure}[!h]
\centering
\includegraphics[width=\textwidth]{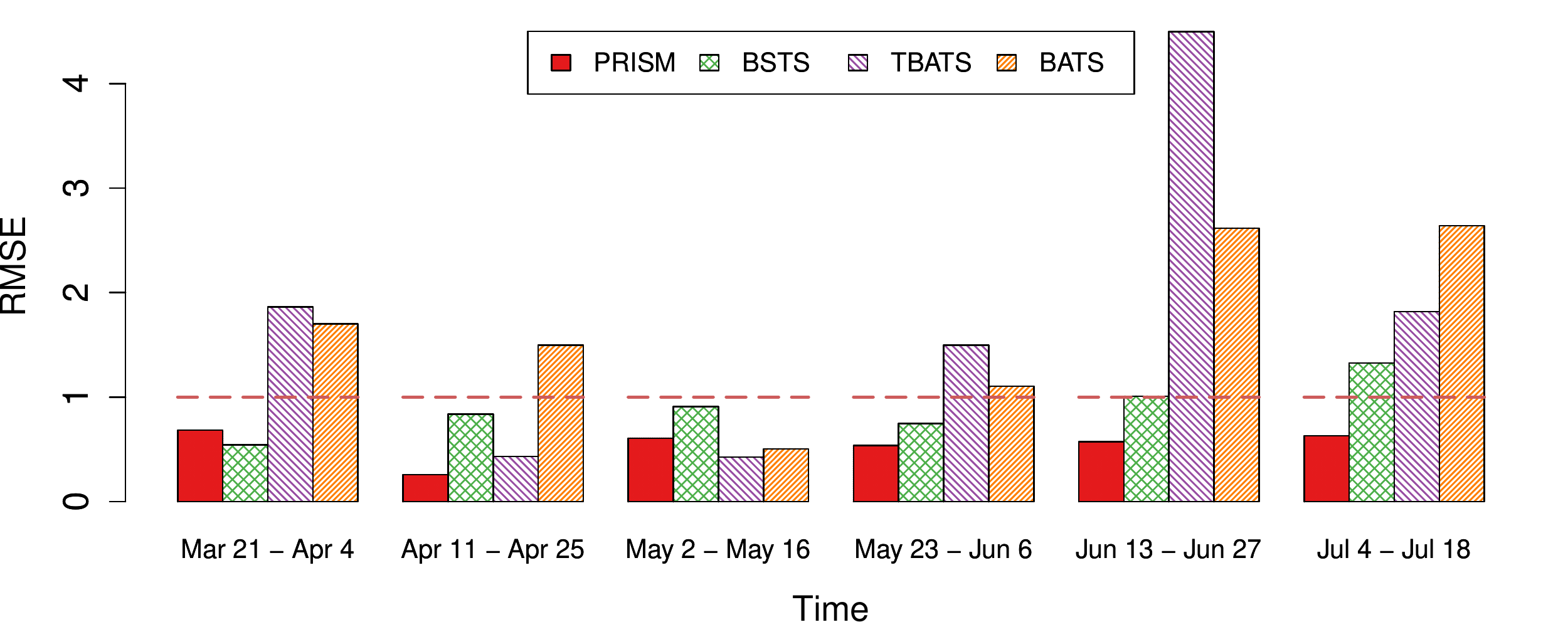}
\caption{Nowcasting performance of different methods during the COVID-19 pandemic period (March 21, 2020 to July 18, 2020). RMSE is measured relative to the error of the naive method and evaluated in three-week windows; a value above 1 indicates that the method performs worse than the naive method in that time period.}
\label{rmse_covid}
\end{figure}

Table \ref{overall_covid} and Figure \ref{rmse_covid} summarize the performance of PRISM's real-time nowcasting of the weekly unemployment claims in comparison with the other methods during the COVID pandemic period from March 21, 2020 to July 18, 2020. For close examination of the different methods, we break down the entire period into 3-week windows. During the first three weeks when the COVID-19 shutdown triggered the sudden and drastic rise in unemployment claims, both PRISM and BSTS picked up the signal rather quickly due to the input from the real-time Google Trends data that track the search of unemployment related query terms. Since April, PRISM began to show its advantage over the other methods as it adapted the predictive model to the ``new regime'', leading the chart in 5 out of 6 evaluation windows. It is worth pointing out that forecasting unemployment initial claims during this period is a very challenging task as the unprecedented huge jump of unemployment claims drastically altered the pattern in the data (including the time-series pattern): we noted that PRISM is the only method that consistently outperforms the naive method throughout this COVID-19 period (the time-series based methods often performed worse than the naive method). This out-of-sample forecasting performance thus indicates the robustness of PRISM to unusual economic shocks and events, giving us more evidence of the model's reliability and accuracy. Further evaluation of PRISM and the benchmarks for longer-horizon predictions is presented in Supplementary Material A16, where PRISM also shows advantage in near-future predictions.

\begin{table}[!h]
\footnotesize
\begin{center}
\begin{adjustbox}{center}
\begin{tabular}{lrrrrrrr}
  \hline

 & Mar 21-Apr 4 & Apr 11-Apr 25 & May 2-May 16 & May 23-Jun 6 & Jun 13-Jun 27 & Jul 4-Jul 18  \\ 
  \hline
    RMSE\\

\quad  PRISM &  0.684 & \textbf{0.257} & 0.607 & \textbf{0.538} & \textbf{0.574} & \textbf{0.629} \\ 
\quad   BSTS & \textbf{0.543} & 0.838 & 0.910 & 0.749 & 1.009 & 1.327 \\ 
\quad   BATS & 1.701 & 1.498 & 0.505 & 1.104 & 2.617 & 2.640 \\ 
\quad   TBATS & 1.862 & 0.432 & \textbf{0.425} & 1.497 & 4.496 & 1.817 \\ 
\quad  naive & 1 (2362454) & 1 (932295) & 1 (488192) & 1 (232074) & 1 (59761) & 1 (105392) \\ 
  MAE\\

\quad  PRISM & 0.784 & \textbf{0.215} & 0.462 & \textbf{0.445} & \textbf{0.711} & \textbf{0.488} \\ 
\quad   BSTS & \textbf{0.539} & 0.863 & 0.879 & 0.700 & 1.092 & 1.460 \\ 
\quad   BATS & 1.952 & 1.514 & 0.553 & 0.896 & 3.236 & 2.762 \\ 
\quad   TBATS & 2.136 & 0.363 & \textbf{0.453} & 1.651 & 5.940 & 1.818 \\ 
 \quad    naive & 1 (1986663) & 1 (898656) & 1 (444600) & 1 (206791) & 1 (44883) & 1 (95054) \\ 
     \hline
\end{tabular}
\end{adjustbox}
\caption{Performance of PRISM and benchmark methods during the COVID-19 pandemic period for real-time nowcasting. Evaluation period is broken down to 3-week windows. RMSE and MAE here are relative to the error of naive method; that is, the number reported is the ratio of the error of a given method to that of the naive method; the absolute RMSE and MAE of the naive method are reported in the parentheses. The boldface indicates the best performer for each forecasting horizon and each accuracy metric.}\label{overall_covid}
\end{center}
\end{table}

\section{Discussion}
The wide availability of data generated from the Internet offers great potential for predictive analysis and decision making.  Our study on using Internet search data to forecast unemployment initial claims illustrates one such potential. The arrival of new data (sometimes in new forms) requires new methodology to analyze and utilize them. PRISM is an example where traditional statistical models are brought together with more recent statistical tools, such as $L_1$ regularization and dynamic training. 
 
In this article we focus on using Internet search data to forecast unemployment initial claims weeks into the future. We introduced a novel method PRISM for forecasting time series with strong seasonality. PRISM is semi-parametric and can be generally applied with or without exogenous variables. PRISM is motivated from a general state-space formulation that contains a variety of widely used time series models as special cases. The two stages of PRISM are easy to implement. The numerical evaluation shows that PRISM outperforms all alternatives in forecasting unemployment initial claim data for the time period of $2007-2019$. We believe that the accurate and robust forecasts by PRISM would greatly benefit the public and the private sectors to assess and gauge the economic trends.

PRISM also demonstrates stable adaptability to unusual economic shocks such as the 2008-2009 financial crisis and the 2020 COVID-19 pandemic shutdown. The out-performance of PRISM relative to other methods are robust during long periods of economic expansion and during short periods of economic recession. In particular, during the 2008-2009 financial crisis and the COVID-19 pandemic period, we found that the real-time data from Google enables PRISM to quickly pick up the signal and the changes in data patterns and to provide insight on real-time and near-future economic trends. This gives us confidence that the unemployment forecasts given by PRISM would provide government agencies with much-needed 
information to react promptly and make well-informed decisions in the face of future economic and financial shocks.

The predictive power and advantage of PRISM mainly come from the following features: (1) dynamic model training based on a rolling window to account for changes in people's search pattern and changes in the relationship between Google search information and the targeted economic activity/index; (2) utilization of $L_1$ penalty to select the most relevant predictors and to filter out noisy and redundant information; (3) combination of non-parametric seasonality decomposition and penalized regression for greater flexibility and adaptability; (4) incorporation of real-time Google search information from multiple related query terms to enhance prediction accuracy and robustness.

Although this article focuses on forecasting unemployment initial claims, PRISM can be generally used to forecast time series with complex seasonal patterns. 
The semi-parametric approach of PRISM covers a wider range of time series models than traditional methods, as PRISM transforms the inference of a complicated class of state-space models into penalized regression of linear predictive models.
Furthermore, dynamically fitting the predictive equations of PRISM addresses the time-varying relationship between the exogenous variables and the underlying time series. One interesting question for future study is to explore if we can extend PRISM to forecasting unemployment indicators in more specified industries such as construction, manufacturing, transportation, finance, and government or to forecasting other unemployment indicators such as non-farm payrolls. Another direction for future study is to extend PRISM to predict unemployment indicators for different ethnic or demographic groups. Furthermore, it would also be of great future interests to see if PRISM can contribute to forecasting future breaks and macro-economic cycles.

We conclude this article with a few remarks on the detailed implementation of the PRISM method. We used real-time Internet search data from Google Trends, which provides publicly available data through subsampling and renormalization: a data set undergoes subsampling (Google draws a small sample from its raw search data for a search query term) and renormalization (after the sampling, Google normalizes and rounds up the search volumes so that they become integers between 0 and 100 for each search query term) when downloaded.
Due to the subsampling and renormalization, the search term volumes are noisy and variable \citep{yang2015accurate}. The $L_1$ regularization adopted in PRISM has shown advantage in extracting the signals and reducing redundant information from Google search data (see Supplementary Material A9 and A11). 
Furthermore, the dynamic training with rolling window accounts for changes in search engine algorithms, people's search patterns, economic trends and other patterns that change over time \citep{burkom2007automated, ning2019accurate}. This is also evident in Figure A11, where each of the 25 candidate search terms has distinct patterns coming in and out of the dynamically fitted model throughout the time. The discount factor adopted also gives more weights on more recent data to capture more recent economic trends and Google search changes, which is similar to the data tapering idea proposed by \citep{dahlhaus1988small, dahlhaus1997fitting} for improved performance in locally stationary time series. 
Our empirical analysis supports the effectiveness of the rolling window and discount factor (Supplementary Material A6 and A7). Alternative frameworks for inferring time series models with state-space structures include the dynamic linear model (DLM) \citep{shumway2017time}.

One limitation with PRISM arises from the data availability from Google Trends. The public available tool only provides up to 5-year range of weekly search data per download. Access to data in higher resolution and longer time span requires Google's permission to use its nonpublic API. Furthermore, in each downloaded batch, the search volume data are normalized by Google to the scale from 0 to 100 based on the queried search term and specific data range of that download. Thus, to avoid the variability due to the normalization and to ensure the consistency of results, we kept both the model training and the corresponding prediction within the same (downloaded) set of 5-year data. Furthermore, Google search data may not reflect the entire population of unemployed people, especially those who would not search online for employment information. Therefore, what we utilized is essentially the association between the search volume of related search terms and our target of unemployment claims for PRISM's prediction.

The R package \verb|PRISM.forecast| that implements the PRISM method is available on CRAN at \url{https://CRAN.R-project.org/package=PRISM.forecast}. We also made the code available at \url{https://github.com/ryanddyi/prism}.

\section*{Acknowledgment}
S.C.K.'s research is supported in part by NSF grant DMS-1810914.

\baselineskip12pt

\bibpunct{(}{)}{;}{a}{}{,} 
\bibliographystyle{jasa}
\bibliography{reference}

\newpage
\setcounter{page}{1}
\begin{center}
\bf \huge 
Supplementary Material
\end{center}

\bigskip 

\setcounter{equation}{0}
\setcounter{section}{0}
\setcounter{figure}{0}
\setcounter{table}{0}

\renewcommand {\thefigure} {A\arabic{figure}}
\renewcommand {\thetable} {A\arabic{table}}
\renewcommand {\theequation} {A\arabic{section}.\arabic{equation}}
\renewcommand {\thesection} {A\arabic{section}}


Details of the methodology, derivation and performance of PRISM are presented as follows. First, the exact search query terms used in our study, which were identified from Google Trends, are presented. Second, the general state-space formulation that motivates PRISM is presented together with a few widely used special cases. Third, the predictive distribution for PRISM forecasting, together with the mathematical proof, is described in detail. Fourth, the robustness of PRISM to the choice of the seasonal decomposition method, to the choice of the discount factor, to the choice of the length of the training window, and to the choice of the number of observations for the seasonal decomposition is presented. 
Fifth, the effect of regularization in PRISM is studied. Sixth, we investigate the normality of the residuals, which supports PRISM's predictive interval construction. Seventh, the fitted coefficients of PRISM are discussed. Eighth, we further compare PRISM with two additional benchmark methods. Ninth, one additional performance metric, CSSED, is studied for comparing PRISM with the alternative methods. Tenth, year-by-year forecasting performance of different methods in longer forecasting horizons (1-week, 2-week, and 3-week ahead) are reported. Eleventh, the predictive intervals of PRISM for longer horizon forecasting (1-week, 2-week, and 3-week ahead) are presented.
Twelfth, we report the results of different methods for forecasting longer horizons during the COVID-19 pandemic period.

\section{Internet Search Data from Google} \label{supp_data}

\begin{figure*}[!htbp]
\centering
\includegraphics[width=\textwidth]{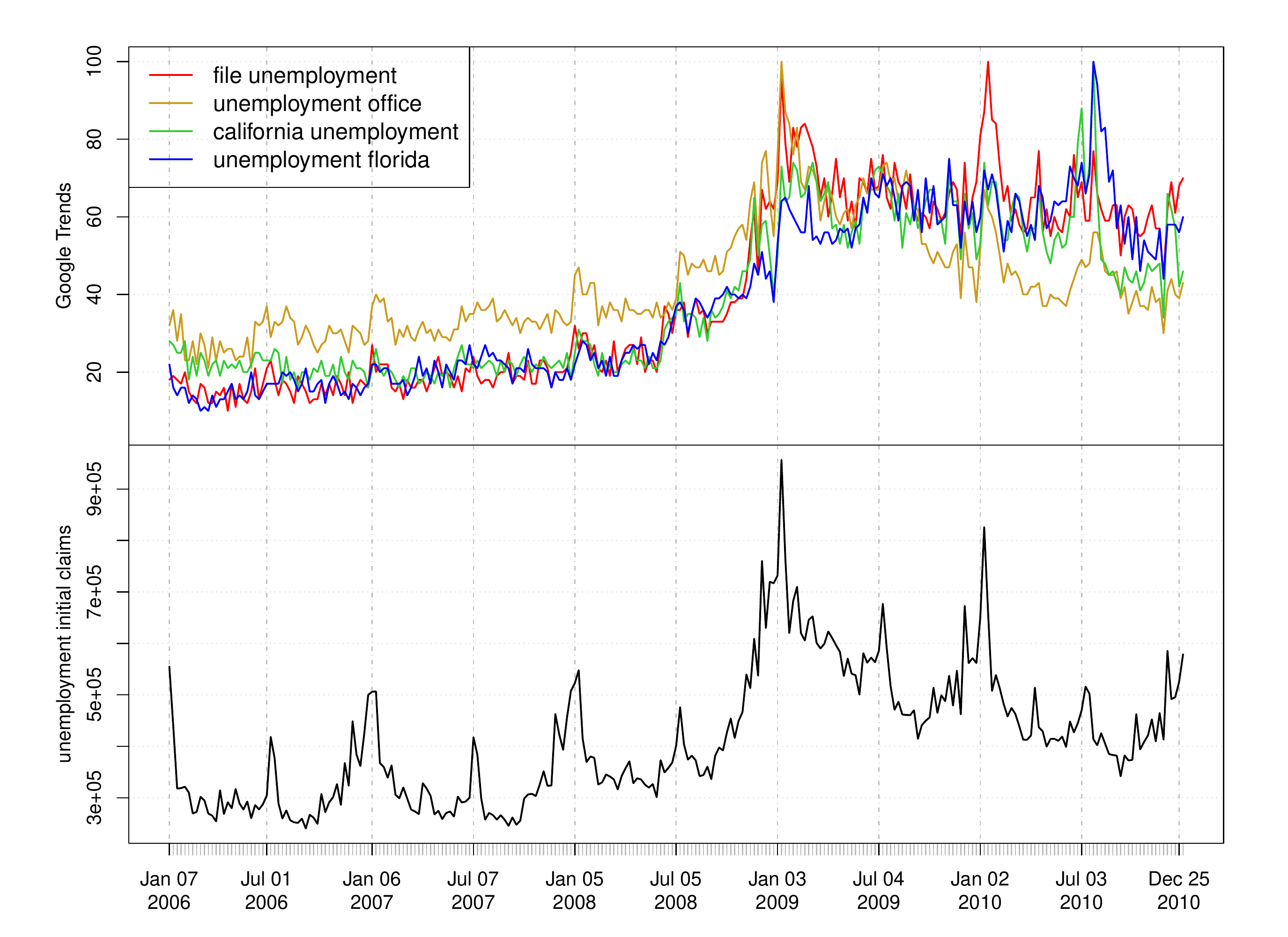}
\caption{The upper panel shows the Google Trends data of four unemployment related search queries in 2006-2010. The lower panel shows the weekly unemployment initial claims data in the same time period.}
\label{GT_plot}
\end{figure*}

The real-time Internet search data we used were from Google Trends (\url{www.google.com/trends}). The search query terms that we used in our study were also identified from the Google Trends tool. One feature of Google Trends is that, in addition to the time series of a specific term (or a general topic), it also returns the top query terms that are most highly correlated with the specific term. In our study, we used a list of top 25 Google search terms that are the most highly correlated with the term ``unemployment''.  Table \ref{table_search_terms} lists these 25 terms, which were generated by Google Trends on January 11, 2018. Figure \ref{GT_plot}, the upper panel, illustrates the Google Trend time series of several search query terms in a 5-year span. Comparing these time series to the lower panel of Figure \ref{GT_plot}, which shows the unemployment initial claims in the same time period, it is evident that the former provides noisy signal about the latter. On the Google Trends site, the weekly data are available for at most a 5-year span in a query, and it would be automatically transformed to monthly data if one asks for more than 5 years. To model and forecast the weekly unemployment claims for the entire period of 2007-2016, we downloaded separate weekly data sets from Google Trends, covering 2004-2008, 2006-2010, 2008-2012, 2010-2014 and 2012-2016, respectively. 

\begin{figure*}[!htbp]
\centering
\includegraphics[width=\textwidth]{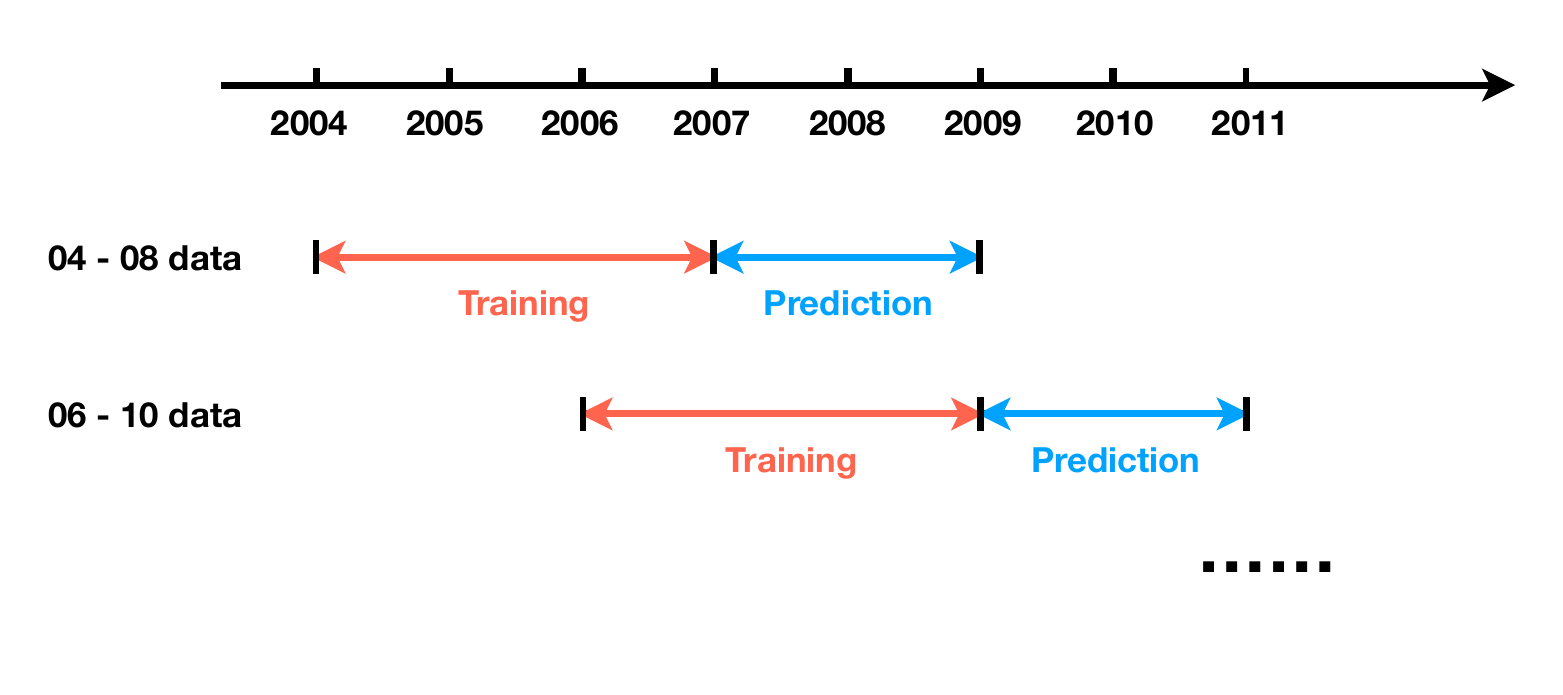}
\caption{Illustration of the rolling-window prediction scheme based on 5-year span of Google Trends data.}
\label{GTdata}
\end{figure*}

To avoid the variability due to the normalization and to ensure the coherence of the model, for each week, we kept both the training data and the data used for prediction within the same 5-year span of data downloaded. For each search term, we downloaded Google Trends data based on the same 5-year range and 
the multivariate $\boldsymbol{x}_t$. Then we trained the model and made predictions based on a rolling window of 3 years. So for the same set of data downloaded with the range of 2004-2008, we are able to make predictions for weeks in 2007-2008; similarly, the data covering 2006-2010 will give predictions for weeks in 2009-2010, etc. See Figure \ref{GTdata} for an illustration.

\begin{table}[h]
\caption{Top 25 nationwide search query terms associated with the term ``unemployment'' generated by Google Trends as of January 11, 2018.}
\scriptsize
\centering
\begin{tabular}{lll}
  \hline
  unemployment & unemployment benefits & unemployment rate \\ 
  unemployment office & pa unemployment & claim unemployment \\ 
  ny unemployment & nys unemployment & ohio unemployment \\ 
  unemployment florida & unemployment extension & texas unemployment \\ 
  nj unemployment & unemployment number & file unemployment \\ 
  unemployment insurance & california unemployment & unemployed \\ 
  unemployment oregon & new york unemployment & indiana unemployment \\ 
  unemployment washington & unemployment wisconsin & unemployment online \\ 
  unemployment login &  & \\ 
 
   \hline
\end{tabular}
\label{table_search_terms}
\end{table}

\section{Our State-Space Formulation and Special Cases} \label{supp:statespace}

Our state-space model, Eq. (\ref{uni}) in the main text, contains a variety of widely used time series models, including structural time series models and additive innovation state-space models. Under this general formulation, a specific parametric model can be obtained by specifying the state-space models for $z_t$ and $\gamma_t$ along with their dependence structure $\bm{H}$. 
\subsection{Special case 1: AR model with seasonal pattern}

The following AR model with seasonal pattern is a special case: modeling $z_t$ as an autoregressive process with lag $N$ and assuming a dummy variable formulation with period $S$ for the seasonal component $\gamma_t$: 
\begin{equation}
\begin{aligned}
y_t &= z_t+\gamma_t,\\
z_t &= \mu_z+\sum_{j=1}^{N}\alpha_j z_{t-j}+\eta_t,\quad \eta_t\overset{\text{iid}}{\sim}  \mathcal{N}(0,\sigma_\eta^2)\\
\gamma_t &= -\sum_{j=1}^{S-1}\gamma_{t-j}+\omega_t,\quad \omega_t\overset{\text{iid}}{\sim} \mathcal{N}(0,\sigma_\omega^2)
\end{aligned}
\label{sar}
\end{equation}
The dummy variable model for the seasonal component implies that sum of the seasonal components over the $S$ periods, $\sum_{j=0}^{S-1}\gamma_{t-j}$, has mean zero and variance $\sigma_\omega^2$. 
In the model (\ref{sar}), each time series block of $\{z_{(t-N+1):t}\}_{t\geq N}$ and $\{\gamma_{(t-S+2):t}\}_{t\geq (S-1)}$ evolves as a Markov Chain. Under our general state-space model, Eq. (\ref{uni}) in the main text, if we set $\bm{h}_t = (1, z_t,z_{t-1},\ldots,z_{t-N+1})$ and $\bm{s}_t=(\gamma_t, \gamma_{t-1}, \ldots, \gamma_{t-S+2})$, then it reduces to the model (\ref{sar}).

\subsection{Special case 2: structural time series models}
The basic structural model assumes that a univariate time series is the sum of trend, seasonal and irregular components, each of which follows an independent stochastic process \citep{harvey1989forecasting}. The model is 
\begin{equation}
y_t = \mu_t + \gamma_t + \epsilon_t,\qquad \epsilon_t \overset{iid}{\sim} \mathcal{N}(0, \sigma_\epsilon^2),
\label{bsm1}
\end{equation}
where $\mu_t$ is the trend component, and $\gamma_t $ and $\epsilon_t$ are the seasonal and irregular components, respectively. 
The trend is often specified by a local level model
\begin{subequations}
\begin{eqnarray}
\mu_t &=& \mu_{t-1} + \delta_t + \eta_t,\qquad \eta_t \overset{iid}{\sim} \mathcal{N}(0, \sigma_\eta^2),\\
\delta_t &=& \delta_{t-1} + \zeta_t,\qquad \zeta_t \overset{iid}{\sim} \mathcal{N}(0, \sigma_\zeta^2),
\end{eqnarray}
\label{bsm2}
\end{subequations}
where $\mu_t$ is the level and $\delta_t$ is the slope. $\eta_t$ and $\zeta_t$ are assumed mutually independent. For time series with $S$ periods, the seasonal component can be specified through the seasonal dummy variable model
\begin{equation}
\gamma_t = -\sum_{j=1}^{S-1}\gamma_{t-j}+\omega_t,\qquad \omega_t \overset{iid}{\sim} \mathcal{N}(0, \sigma_\omega^2).\\
\label{bsm3}
\end{equation}
which is the same as the seasonal component in the seasonal AR model \eqref{sar}. 
Alternatively, the seasonal pattern can be modeled by a set of trigonometric terms at seasonal frequencies $\lambda_j = 2\pi j/S$ \citep{harvey1989forecasting}: 
\begin{subequations}
\begin{eqnarray}
\gamma_t &=& \sum_{j=1}^{[S/2]}\gamma_{j,t},\\
\left(\begin{array}{c}
\gamma_{j,t}\\
\gamma_{j,t}^{*}
\end{array}\right) &=& \left(\begin{array}{cc}
\cos\lambda_{j} & \sin\lambda_{j}\\
-\sin\lambda_{j} & \cos\lambda_{j}
\end{array}\right)\left(\begin{array}{c}
\gamma_{j,t-1}\\
\gamma_{j,t-1}^{*}
\end{array}\right)+\left(\begin{array}{c}
\omega_{j,t}\\
\omega_{j,t}^{*}
\end{array}\right),
\end{eqnarray}
\label{trigo}
\end{subequations}
where $\omega_{j,t}$ and $\omega_{j,t}^{*}$, $j=1,\ldots,[S/2]$, are independent and normally distributed with common variance $\sigma_\omega^2$. 

Under our general state-space model, Eq. (\ref{uni}) of the main text, if we take $z_t = \mu_t + \epsilon_t$ and $\bm{h}_t = (\mu_t, \delta_t)$, then it specializes to structural time series models. In particular, for the dummy variable seasonality of \eqref{bsm3}, $\bm{s}_t$ in the general model corresponds to $\bm{s}_t = (\gamma_t, \gamma_{t-1}, \ldots, \gamma_{t-S+2})$; and for the trigonometric seasonality of \eqref{trigo}, $\bm{s}_t$ in the general model corresponds to $\bm{s}_t = (\gamma_{1,t},\ldots,\gamma_{[S/2],t}, \gamma^{*}_{1,t},\ldots,\gamma^{*}_{[S/2],t})$. 

\subsection{Special case 3: additive innovations state-space models}

An alternative to structural time series models, which have multiple sources of error, innovation state-space model \citep{aoki1987state}, where the same error term appears in each equation, is also popular. These innovation state-space models underlie exponential smoothing methods, which are widely used in time series forecasting and have been proven optimal under many specifications of the innovation state-space model \citep{ord1997estimation,hyndman2008forecasting}. 
Among exponential smoothing methods, Holt-Winters' method \citep{holt1957forecasting, winters1960forecasting} is developed to capture both trend and seasonality, and it postulates a model specification similar to the basic structural model \eqref{bsm1}- \eqref{bsm3}. In particular, Holt-Winters' additive method is
\begin{subequations}
\begin{eqnarray}
y_t &=& \mu_{t-1} + \delta_{t-1} + \gamma_{t-S} + \epsilon_t,\\
\mu_t &=& \mu_{t-1} + \delta_{t-1} + \alpha\epsilon_t,\\
\delta_t &=& \delta_{t-1} + \beta\epsilon_t,\\
\gamma_t &=& \gamma_{t-S} + \omega\epsilon_t,
\end{eqnarray}
\label{etsaaa}
\end{subequations}
where the components $\mu_t$, $\delta_t$ and $\gamma_t$ represent level, slope and seasonal components of time series, and $\epsilon_t \overset{iid}{\sim} \mathcal{N}(0, \sigma^2)$ is the only source of error. 
Since Eq. (\ref{etsaaa}a) can be rewritten as 
\begin{equation*}
y_t = \mu_t + \gamma_t + (1-\alpha - \omega)\epsilon_t,
\label{ets_rw}	
\end{equation*}
we observe that model (\ref{etsaaa}) is special case of our general formulation with $z_t = \mu_t + (1-\alpha-\omega)\epsilon_t$, $\bm{h}_t = (\mu_t, \delta_t)$ and $\bm{s}_t = \left(\gamma_t, \gamma_{t-1}, \ldots, \gamma_{t-S+1}\right)$ in Eq. (\ref{uni}) of the main text.
The Holt-Winters model is among a collection of innovation state-space models that are summarized in  \cite{hyndman2008forecasting} using the triplet (E, T, S), which represents model specification for the three components: error, trend and seasonality. For instance, \eqref{etsaaa} is also referred to as local additive seasonal model or ETS(A,A,A), where A stands for additive. Our general state-space formulation, Eq. (\ref{uni}) in the main text, also incorporates many useful model extensions as special cases, including the damped trend \citep{gardner1985forecasting} and multiple seasonal patterns \citep{gould2008forecasting, de2011forecasting}. For example, the damped trend double seasonal model extends model \eqref{etsaaa} to include a factor $\phi\in [0,1)$ and a second seasonal component as follows:
\begin{equation}
\begin{aligned}
y_t &= \mu_{t-1} + \phi\delta_{t-1} + \gamma^{(1)}_{t-S_1} + \gamma^{(2)}_{t-S_2} + \epsilon_t,\\
\mu_t &= \mu_{t-1} + \phi\delta_{t-1} + \alpha\epsilon_t,\\
\delta_t &= \phi\delta_{t-1} + \beta\epsilon_t, \\
\gamma_t^{(1)} &= \gamma^{(1)}_{t-S_1} + \omega_1\epsilon_t, \\
\gamma_t^{(2)} &= \gamma^{(1)}_{t-S_2} + \omega_2\epsilon_t.
\end{aligned}
\label{extend}
\end{equation}
Our general model contains this extended model, where $z_t = \mu_t + (1-\alpha-\omega_1-\omega_2)\epsilon_t$, $\gamma_t = \gamma^{(1)}_t + \gamma^{(2)}_t$, $\bm{h}_t = \left(\mu_t, \delta_t\right)$ and $\bm{s}_t = (\gamma^{(1)}_t, \ldots, \gamma^{(1)}_{t-S_1+1}, \gamma^{(2)}_t, \ldots, \gamma^{(2)}_{t-S_2+1})$.

\subsection{Motivation of the general formulation}

The motivation of our general state-space formulation is to collectively consider all possible models under it and to \emph{semi-parametrically} obtain the prediction under this large class of models. 
In comparison, traditional time series studies often rely on parameter estimation of specified models such as those highlighted in the previous subsections. 
For instance, exponential smoothing is tailored for computing the likelihood and obtaining maximum likelihood estimates of the innovation state-space models. 
For other parametric models with multiple sources of error, their likelihood might be evaluated by the Kalman filter, but the parameter estimation can be difficult in many cases. In the traditional parametric times series model setting, model selections are often applied by optimizing certain selection criteria (e.g. AIC or BIC), but when the class of models under consideration become really large such as Eq. (\ref{uni}) of the main text, traditional model selection methods encounter serious challenges (as they lack scalability) to operate on such a wide range of models. As a consequence, traditional parametric time series models often consider a much smaller collection of models compared to Eq. (\ref{uni}) of the main text. The cost of focusing on a small class of models is that the forecasting accuracy can substantially suffer as the risk of model misspecification is high.

To relieve these challenges and improve the performance of forecasting, we use our general state-space formulation to motivate a semi-parametric method for forecasting time series. We derive and study a linear predictive model that is coherent with all possible models under Eq. (\ref{uni}) of the main text. With forecasting as our main goal, we essentially transform the question from the inference of a complicated class of state-space models into penalized regression and forecasting based on a linear prediction formulation. 

\section{Predictive Distributions for Forecasting}

Under our general state-space model -- Eq. (\ref{uni}) and Eq. (\ref{x_multi}) of the main text -- given the historical data $\{y_{1:(t-1)}\}$ and contemporaneous exogenous time series $\{\bm{x}_{t_0:t}\}$, the predictive distribution for forecasting $y_{t+l}$ ($l\geq 0$) at time $t$ would be $p(y_{t+l}\mid y_{1:(t-1)}, \bm{x}_{t_0:t})$. In PRISM, we consider the predictive distribution of $y_t$ by further conditioning on the latent seasonal component $\{\gamma_t\}$:
\begin{equation}
p(y_{t+l}\mid y_{1:(t-1)}, \gamma_{1:(t-1)}, \bm{x}_{t_0:t}).
\label{full_con}
\end{equation}
Note that since $z_t=y_t-\gamma_t$ for all $t$, $z_{1:(t-1)}$ is known given $y_{1:(t-1)}$ and $\gamma_{1:(t-1)}$. The advantage of working on \eqref{full_con} is that we can establish a universal representation of the predictive distribution as given by the next proposition.  
\begin{prop}
Under our model --- Eq. (\ref{uni}) and (\ref{x_multi}) of the main text --- $y_{t+l}$ ($l\geq 0$) conditioning on $\{ z_{1:(t-1)}, \gamma_{1:(t-1)}, \bm{x}_{t_0:t}\}$ follows a normal distribution with the conditional mean $E(y_{t+l}\mid z_{1:(t-1)}, \gamma_{1:(t-1)}, \bm{x}_{t_0:t})$ linear in $z_{1:(t-1)}$, $\gamma_{1:(t-1)}$ and $\bm{x}_t$.
\label{prop1}
\end{prop}

As a partial result that leads to Proposition \ref{prop1}, we have, without the exogenous variables, the conditional distribution of $y_{t+l}\mid z_{1:(t-1)}, \gamma_{1:(t-1)}$ is normal with mean linear in $z_{1:(t-1)}$ and $\gamma_{1:(t-1)}$.

Based on Proposition \ref{prop1}, we can represent $p\left(y_{t+l}\mid z_{1:(t-1)}, \gamma_{1:(t-1)}, \bm{x}_{t_0:t}\right)$ as
\begin{equation}
y_{t+l}=\mu_t^{(l)}+\sum_{j=1}^{t-1}\alpha_{j,t}^{(l)} z_{t-j}+\sum_{j=1}^{t-1}\delta_{j,t}^{(l)}\gamma_{t-j}+\sum_{i=1}^{p}\beta_{i,t}^{(l)}x_{i,t}+\epsilon_t,\quad \epsilon_t\overset{iid}{\sim}  \mathcal{N}(0,\sigma_{t,l}^2),
\label{eq_pred_forecast}
\end{equation}
where $\mu_t^{(l)}$, $\alpha_{j,t}^{(l)}$, $\delta_{j,t}^{(l)}$, $\beta_{i,t}^{(l)}$ and $\sigma_{t,l}^2$ are fixed but unknown constants that are determined by original parameters $\bm{\theta}$ and the initial values of the state vectors. 

\section{Proof of Proposition \ref{prop1}}

We first prove the case for $l=0$.
Let $\bm{r}_t = (z_t,\gamma_t)'$. Since $y_t=z_t+\gamma_t$ for all $t$, $y_t\mid y_{1:(t-1)}, \gamma_{1:(t-1)}$ is equivalent to $z_t+\gamma_t \mid z_{1:(t-1)}, \gamma_{1:(t-1)}$. By treating $\bm{r}_t$ as a 2-dimensional observable and $\bm{\alpha}_t=(\bm{h}_t', \bm{s}_t')'$ as the state vector, we can rewrite Eq. (\ref{uni}) of the main text as
\begin{equation}
\begin{aligned}
\bm{r}_t &= \bm{\Phi}'\bm{\alpha}_t+\bm{\xi}_t\\
\bm{\alpha}_{t} &= \bm{\Lambda}\bm{\alpha}_{t-1}+\bm{\tau}_t
\end{aligned}
\label{uni_rw}
\end{equation}
where $\bm{\Phi}=\left[\begin{array}{cc}
\bm{w} & \bm{0}\\
\bm{0} & \bm{v}
\end{array}\right]$, $\bm{\Lambda}=\left[\begin{array}{cc}
\bm{F} & \bm{O}\\
\bm{O} & \bm{P}
\end{array}\right]$ and $(\bm{\xi}_t', \bm{\tau}_t')' \overset{iid}{\sim} \mathcal{N}(\bm{0}, \bm{H})$. 

Denote $\bm{r}_{1:t} = (\bm{r}'_1,\ldots, \bm{r}'_t)'$ and  $\bm{\alpha}_{1:t} = (\bm{\alpha}'_1,\ldots, \bm{\alpha}'_t)'$. According to the property of Gaussian linear state-space model, $\bm{r}_{1:t}$ and $\bm{\alpha}_{1:t}$ jointly follows a multivariate normal distribution. Therefore, the sub-vector $\bm{r}_{1:t}$ is also normal, and $\bm{r}_{t} \mid \bm{r}_{1:(t-1)}$ follows a bivariate normal distribution with mean linear in $\bm{r}_{1:(t-1)}$, i.e.
\begin{equation}
\bm{r}_t \mid \bm{r}_{1:(t-1)} \sim \mathcal{N}(\bm{\Gamma}_t\bm{r}_{1:(t-1)}, \bm{\Sigma}_t),
\label{eq_pf1}
\end{equation}
where $\bm{\Gamma}_t$ and $\bm{\Sigma}_t$ are determined by $\bm{\Phi}$, $\bm{\Lambda}$ and $\bm{H}$. For any given parameters, the above distribution can be numerically evaluated through Kalman filter. Here, we focus only on the general analytical formulation. Following \eqref{eq_pf1} and $y_t = \bm{1}'\bm{r}_t$, we have $$y_t \mid \bm{r}_{1:(t-1)} \sim \mathcal{N}(\bm{1}'\bm{\Gamma}_t\bm{r}_{1:(t-1)}, \bm{1}'\bm{\Sigma}_t\bm{1}).$$
Thus, given $\bm{r}_{1:(t-1)}$, or equivalently $z_{1:(t-1)}$ and $\gamma_{1:(t-1)}$, $y_t$ has a univariate normal distribution with mean linear in $z_{1:(t-1)}$ and $\gamma_{1:(t-1)}$. 

When taking the exogenous variable $\bm{x}_t$ into account, we have $p(y_t\mid \bm{r}_{1:(t-1)}, \bm{x}_{t_0:t})\propto p(\bm{x}_t\mid y_t)p(y_t\mid \bm{r}_{1:(t-1)})$. Since 
\begin{equation}
p(\bm{x}_t\mid y_t)\propto \exp\left(-\frac{1}{2}(\bm{x}_t-\bm{\mu}_x-y_t\bm{\beta})'\bm{Q}^{-1}(\bm{x}_t-\bm{\mu}_x-y_t\bm{\beta})\right),
\label{xgiveny}
\end{equation}
it follows that 
\begin{eqnarray*}
& & p(y_t\mid \bm{r}_{1:(t-1)}, \bm{x}_{t_0:t}) \\
&\propto& \exp\left(-\frac{1}{2}(\bm{x}_t-\bm{\mu}_x-y_t\bm{\beta})'\bm{Q}^{-1}(\bm{x}_t-\bm{\mu}_x-y_t\bm{\beta})-\frac{1}{2}(\bm{1}'\bm{\Sigma}_t\bm{1})^{-1}(y_t-\bm{1}'\bm{\Gamma}_t\bm{r}_{1:(t-1)})^2\right).
\end{eqnarray*}
Hence, $y_t\mid \bm{r}_{1:(t-1)}, \bm{x}_{t_0:t}$ is normal, since the above equation is an exponential function of a quadratic form of $y_t$. By reorganizing the terms, we have 
\begin{equation*}
E\left(y_t\mid \bm{r}_{1:(t-1)}, \bm{x}_{t_0:t}\right) = \left(\left(\bm{1}'\bm{\Sigma}_{t}\bm{1}\right)^{-1}+\bm{\beta}'\bm{Q}^{-1}\bm{\beta}\right)^{-1}\left(\left(\bm{1}'\bm{\Sigma}_{t}\bm{1}\right)^{-1}\bm{1}'\bm{\Gamma}_{t}\bm{r}_{1:(t-1)} +\bm{\beta}'\bm{Q}^{-1}(\bm{x}_t-\bm{\mu}_x) \vphantom{\frac{1}{\sigma^2}}\right)	
\end{equation*}
and 
\begin{equation*}
Var\left(y_t\mid \bm{r}_{1:(t-1)}, \bm{x}_{t_0:t}\right) = \left(\left(\bm{1}'\bm{\Sigma}_{t}\bm{1}\right)^{-1}+\bm{\beta}'\bm{Q}^{-1}\bm{\beta}\right)^{-1}.	
\end{equation*}
Therefore, $y_t\mid z_{1:(t-1)}, \gamma_{1:(t-1)}, \bm{x}_{t_0:t}$ has a normal distribution with mean linear in $z_{1:(t-1)}$, $\gamma_{1:(t-1)}$ and $\bm{x}_t$.

Next, we prove the case for $l \geq 1$. We consider the following predictive distribution
\begin{eqnarray*}
p\left(y_{t+l}\mid\bm{x}_{t_0:t},\bm{r}_{1:(t-1)}\right)	&\propto & 	\int p\left(y_{t+l},\bm{r}_{t}\mid\bm{x}_{t_0:t},\bm{r}_{1:(t-1)}\right)\text{d}\bm{r}_{t}\\
	&\propto &\int p\left(y_{t+l}\mid\bm{x}_{t_0:t},\bm{r}_{1:t}\right)p\left(\bm{r}_{t}\mid\bm{x}_{t_0:t},\bm{r}_{1:(t-1)}\right)\text{d}\bm{r}_{t}.
\end{eqnarray*}
Since $\bm{x}_{t_0:t}$ is independent of $y_{t+l}$ conditional on $y_{1:t}$, $p(y_{t+l}\mid\bm{x}_{t_0:t},\bm{r}_{1:t}) = p(y_{t+l}\mid\bm{r}_{1:t})$. Similarly, $\bm{x}_{t_0:(t-1)}$ is independent of $\bm{r}_{t}$ conditional on $\bm{r}_{1:(t-1)}$, which implies $p(\bm{r}_{t}\mid\bm{x}_{t_0:t},\bm{r}_{1:(t-1)})=p(\bm{r}_{t}\mid\bm{x}_{t},\bm{r}_{1:(t-1)})$. Note that $p\left(\bm{r}_{t}\mid\bm{x}_{t},\bm{r}_{1:(t-1)}\right)\propto p(\bm{x}_t\mid \bm{r}_t)p(\bm{r}_t\mid \bm{r}_{1:(t-1)})$. Thus, we have 
\begin{eqnarray*}
p\left(y_{t+l},\bm{r}_{t}\mid\bm{x}_{t_0:t},\bm{r}_{1:(t-1)}\right) &\propto & p\left(y_{t+l}\mid\bm{r}_{1:t}\right)p\left(\bm{r}_{t}\mid\bm{x}_{t},\bm{r}_{1:(t-1)}\right)\\
	&\propto & p\left(y_{t+l}\mid \bm{r}_{1:t}\right)p(\bm{x}_t\mid \bm{r}_t)p(\bm{r}_t\mid \bm{r}_{1:(t-1)}).
\end{eqnarray*}
In the first part of the proof, we have learned that $\bm{r}_{1:(t+l)}$ is multivariate normal. Similar to \eqref{eq_pf1}, we can write $\bm{r}_{t+l}\mid\bm{r}_{1:t}$ as
\begin{equation}
\bm{r}_{t+l} \mid \bm{r}_{1:t} \sim \mathcal{N}(\bm{\Gamma}_{t,l}\bm{r}_{1:t}, \bm{\Sigma}_{t,l}),
\label{eq_pf2}
\end{equation}
where $\bm{\Gamma}_{t,l}$ and $\bm{\Sigma}_{t,l}$ are determined by $\bm{\Phi}$, $\bm{\Lambda}$ and $\bm{H}$. Hence, $y_{t+l} \mid \bm{r}_{1:t} \sim \mathcal{N}(\bm{1}'\bm{\Gamma}_{t,l}\bm{r}_{1:t}, \bm{1}'\bm{\Sigma}_{t,l}\bm{1}).$ Combining the above results with \eqref{xgiveny}, we have
\begin{equation}
\begin{split}
p\left(y_{t+l},\bm{r}_{t}\mid\bm{x}_{t_0:t},\bm{r}_{1:(t-1)}\right)\propto \exp\left(-\frac{1}{2}(\bm{1}'\bm{\Sigma}_{t,l}\bm{1})^{-1}\left(y_{t+l}-\bm{1}'\bm{\Gamma}_{t,l}\bm{r}_{1:t}\right)^{2} -\frac{1}{2}(\bm{x}_{t}-\bm{\mu}_{x}-\bm{\beta}\bm{1}'\bm{r}_{t})'\bm{Q}^{-1}\right. \\ 
\left.(\bm{x}_{t}-\bm{\mu}_{x}-\bm{\beta}\bm{1}'\bm{r}_{t}) -\frac{1}{2}(\bm{r}_{t}-\bm{\Gamma}_{t-1,1}\bm{r}_{1:(t-1)})'\bm{\Sigma}_{t-1,1}^{-1}(\bm{r}_{t}-\bm{\Gamma}_{t-1,1}\bm{r}_{1:(t-1)})\right),
\end{split}	
\end{equation}
whose right hand side is an exponential function of a quadratic form of $y_{t+l}$ and $\bm{r}_t$. Hence, $y_{t+l},\bm{r}_{t}\mid\bm{x}_{t_0:t},\bm{r}_{1:(t-1)}$ is multivariate normal. Consequently, the marginal distribution of $y_{t+l} \mid\bm{x}_{t_0:t},\bm{r}_{1:(t-1)}$ is univariate normal. Moreover, the conditional expectation is 
\begin{eqnarray*}
&& E\left(y_{t+l}\mid \bm{x}_{t_0:t},\bm{r}_{1:(t-1)}\right)\\
&=& E\left(E\left(y_{t+l}\mid \bm{r}_{1:t}\right)\mid\bm{x}_{t_0:t},\bm{r}_{1:(t-1)}\right)\\
&=& E\left(\bm{\Gamma}_{t,l}(\bm{r}_{1:(t-1)}',\bm{r}_t')'\mid\bm{x}_{t_0:t},\bm{r}_{1:(t-1)}\right)\\
&=& \bm{\Gamma}_{t,l}\left(\bm{r}_{1:(t-1)}',E(\bm{r}_t\mid\bm{x}_{t},\bm{r}_{1:(t-1)})'\right)',
\end{eqnarray*}
where $E(\bm{r}_t\mid\bm{x}_{t},\bm{r}_{1:(t-1)})$ is linear in $\bm{x}_t$ and $\bm{r}_{1:(t-1)}$. Therefore, $y_{t+l} \mid\bm{x}_{t_0:t},\bm{r}_{1:(t-1)}$ is univariate normal with mean linear in $\bm{x}_t$ and $\bm{r}_{1:(t-1)}$.

\section{Robustness to the Choice of Seasonal Decomposition Method}
\label{robust_decompose}

We compare the performance of PRISM with two seasonal decomposition methods: STL and the classic additive decomposition. Both methods decompose target time series $y_t$ into the trend component $T_t$, the seasonal component $S_t$ and the irregular component $R_t$:
\begin{equation*}
y_t = T_t + S_t + R_t.
\end{equation*}
In the classic additive decomposition, the estimated trend component $\hat{T}_t$ is calculated from the moving average of $\{y_t\}$. The seasonal component $S_t$ for a certain week is assumed to be the same in each period, and is estimated by the average of the detrended value, $y_t - \hat{T}_t$ for the specific week. For example, assuming 52 weeks in a year, the seasonal index for the fifth week is the average of all the detrended fifth week values in the data.

In contrast, STL relies on a sequence of applications of loess smoother to generate the seasonal and trend components. The implementation of STL involves two loops. In the outer loop, robustness weights are assigned to each data point to reduce the effects of outliers, while the inner loop iteratively updates the trend and seasonal components. In the inner loop iteration, the seasonal components are updated using detrended time series similar to classic additive decomposition, and the trend components are calculated by loess smoothing of the deseasonalized time series. 

Both STL and the classic additive seasonal decomposition are options in the R package of PRISM with STL being the default setting. Table \ref{prism_seasonal} describes the performance of PRISM with the two different methods. The numerical result of PRISM is quite robust to the choice of the seasonal decomposition method with STL providing slightly better overall result.

\begin{table}[ht]
\footnotesize
\centering
\caption{The performance of PRISM with two different seasonal decomposition methods: STL and additive decomposition over $2007-2016$.  RMSE and MAE are measured relative to the respective error of the naive method. The boldface highlights the better performance for each metric and forecasting horizon.}

\begin{tabular}{lrrrr}
  \hline
 & real-time & forecast 1 wk & forecast 2 wk & forecast 3 wk \\ 
  \hline
RMSE\\
\qquad additive decomposition &  {0.496} & {0.486} & {0.464} & \textbf{0.467} \\ 
\qquad STL decomposition &
\textbf{0.493} & \textbf{0.483} & \textbf{0.461} & {0.470} \\ 
MAE\\
\qquad additive decomposition & {0.541} & {0.523} & {0.479} & \textbf{0.458} \\ 
\qquad STL decomposition &  \textbf{0.539} & \textbf{0.517} & \textbf{0.476} & {0.460} \\ 
\hline
\end{tabular}
\label{prism_seasonal}
\end{table}

\section{Effect of the Discount Factor}
\label{robust_w}
 
We tested the effect of the discount factor $w$. \cite{lindo1997optimal} suggested setting $w$ between 0.95 and 0.995 for practical applications. Table \ref{prism_w} shows the performance of PRISM with different $w \in  [0.95, 1]$ (note that $w=1$ corresponds to no discount). The performance of PRISM is seen to be quite robust for the different choices of $w$ between $[0.95, 0.995]$. The discount factor $w$ is an option in our R package of PRISM, and the default value is set to be 0.985. We chose $w=0.985$ as it gives the optimal real-time prediction accuracy and close to optimal prediction accuracy in longer forecasting horizons, as reported in Table \ref{prism_w}. We can also see that comparing with no discounting ($w=1$), our default choice of discount factor can provide 1\% to 5\% more error reduction in terms of the relative errors to the naive method. This motivates us to incorporate the discount factor in our model.

\begin{table}[!h]
\footnotesize
\centering
\caption{The performance of PRISM for the discount factor $w \in [0.95, 1]$ over $2007-2016$. RMSE and MAE are measured relative to the respective error of the naive method. The boldface highlights the best performance for each metric and forecasting horizon. }
\begin{tabular}{lrrrr}
  \hline
 & real-time & forecast 1 wk & forecast 2 wk & forecast 3 wk \\ 
  \hline
RMSE\\
\qquad $w=1$ & 0.515 & 0.513 & 0.470 & 0.465 \\ 
\qquad $w=0.995$ & 0.512 & 0.501 & 0.463 & 0.462 \\ 
\qquad $w=0.99$ & 0.495 & 0.491 & \textbf{0.457} & \textbf{0.461} \\ 
\qquad $w=0.985$ & \textbf{0.493} & \textbf{0.483} & 0.461 & 0.470 \\ 
\qquad $w=0.98$ & 0.506 & 0.489 & 0.471 & 0.472 \\ 
\qquad $w=0.975$ & 0.500 & 0.485 & 0.471 & 0.468 \\ 
\qquad $w=0.97$ & 0.513 & 0.488 & 0.475 & 0.495 \\ 
\qquad $w=0.965$ & 0.516 & 0.488 & 0.485 & 0.479 \\ 
\qquad $w=0.96$ & 0.519 & 0.492 & 0.486 & 0.483 \\ 
\qquad $w=0.955$ & 0.525 & 0.502 & 0.502 & 0.497 \\ 
\qquad $w=0.95$ & 0.540 & {0.484} & 0.482 & 0.491 \\ 
   \hline
MAE\\
\qquad $w=1$ & 0.571 & 0.568 & 0.507 & 0.476 \\ 
\qquad $w=0.995$ & 0.556 & 0.546 & 0.487 & 0.467 \\ 
\qquad $w=0.99$ & 0.545 & 0.532 & {0.478} & 0.461 \\ 
\qquad $w=0.985$ & \textbf{0.539} & 0.517 & \textbf{0.476} & \textbf{0.460} \\ 
\qquad $w=0.98$ & 0.543 & 0.517 & 0.484 & \textbf{0.460} \\ 
\qquad $w=0.975$ & 0.540 & \textbf{0.516} & 0.481 & 0.461 \\ 
\qquad $w=0.97$ & 0.549 & 0.519 & 0.488 & 0.482 \\ 
\qquad $w=0.965$ & 0.555 & 0.523 & 0.495 & 0.465 \\ 
\qquad $w=0.96$ & 0.552 & 0.524 & 0.496 & 0.468 \\ 
\qquad $w=0.955$ & 0.563 & 0.533 & 0.511 & 0.477 \\ 
\qquad $w=0.95$ & 0.572 & 0.521 & 0.496 & 0.474 \\ 
\hline
\end{tabular}
\label{prism_w}
\end{table}

\section{The Length of Training Window} \label{Supp:Training}
In this section we conducted an empirical analysis on the choice of the rolling window length in training PRISM. With the limited, 5-year availability of weekly Google search data (more details in Section \ref{sec1.1} and Supplementary Material \ref{supp_data}), we varied the rolling window lengths between 2 and 4 years to keep the fitting and prediction within the same set of downloaded data. Table \ref{tab_window} reports the result for various training window length. We observe that the default 3-year window gives the leading performance. It 
appears that the 3-year window provides a good trade-off between the timeliness of short-term training data and the statistical efficiency from long-term, large-size data. The 3-year rolling window choice is also consistent with the choice in \citet{d2017predictive}. In addition, since Google data only starts in 2004, the choice of 3-year training window enables us to have forecasts from 2007 onward, which is important for evaluating the performance of PRISM (and other methods) during the entire period of the 2008-2009 financial crisis.

\begin{table}[!h]
\footnotesize
\centering
\caption{The RMSE of PRISM under different rolling windows for training over $2007-2016$. RMSE is measured relative to the respective error of the naive method. The boldface highlights the best performance for each forecasting horizon. }\label{tab_window}
\begin{tabular}{lrrrr}
  \hline
 & real-time & 1 week & 2 weeks & 3 weeks \\ 
\hline
PRISM (default 3-year window) & \textbf{0.493} & \textbf{0.484} & \textbf{0.461} & \textbf{0.470} \\ 
PRISM with 2-year window &0.540 & 0.512 & 0.489 & 0.494 \\ 
PRISM with 4-year window &0.502 & 0.494 & 0.467 & 0.476 \\ 
 naive & 1 (50551) & 1 (62227) & 1 (69747) & 1 (73527) \\ 
   \hline
\end{tabular}
\end{table}

\section{Number of Observations for Seasonal Decomposition in Stage 1 of PRISM}
\label{robust_nhist}
For the seasonal decomposition in Stage 1 of PRISM, we used $M=700$ historical initial claim observations as the default choice. A relatively large number of $M$ (for example, more than 10 years of data, $M \geq$ 520) is preferred to ensure stability of the decomposition. We also conducted an experiment on the sensitivity of the choice of $M$. Table \ref{tab_nhist} reports the result, which shows that the performance of PRISM is quite robust to the choice of $M$, and that the default choice of $M=700$ gives rather favorable performance.

\begin{table}[!h]
\footnotesize
\centering
\caption{The performance of PRISM under different choices of $M$, the number of historic initial claim observations for seasonal decomposition in Stage 1 over $2007-2016$. RMSE is measured relative to the respective error of the naive method. The boldface highlights the best performance for each forecasting horizon. }\label{tab_nhist}
\begin{tabular}{lrrrr}
  \hline
 & real-time & 1 week & 2 weeks & 3 weeks \\ 
\hline
PRISM (default $M=700$) & \textbf{0.539} & \textbf{0.518} & \textbf{0.477} & 0.460 \\  
PRISM with $M=600$ & 0.544 & 0.521 & 0.479 & \textbf{0.457} \\ 
PRISM with $M=800$ &  0.544 & 0.521 & 0.479 & 0.464 \\
 naive & 1 (50551) & 1 (62227) & 1 (69747) & 1 (73527) \\ 
   \hline
\end{tabular}
\end{table}

\section{Effect of Regularization} \label{Supp:L2}
We now assess different types of regularization in fitting the coefficients of PRISM, examining the effects of $L_1$, $L_2$ and the elastic net \citep{zou2005regularization} penalty. For $L_2$ regularization, we replaced the $L_1$-norm penalty in \eqref{sargo_l1} with $L_2$ norms (i.e., using the Ridge regression). For the elastic net , the objective function is
\begin{align*}
\frac{1}{N}\sum_{\tau=t-l-N}^{t-l-1} & w^{t-\tau}\Big( y_{\tau+l}-\mu_y-\sum_{j=1}^{K}\alpha_{j} \hat{z}_{\tau-j,\tau}-\sum_{j=1}^{K}\delta_j\hat{\gamma}_{\tau-j,\tau} \nonumber -\sum_{i=1}^{p}\beta_{i}x_{i,\tau}\Big) ^2\\
 &+a\lambda\left(\|\bm{\alpha}\|_{1}+\|\bm{\delta}\|_{1}+\|\bm{\beta}\|_{1}\right)+(1-a)\lambda\left(\|\bm{\alpha}\|_{2}+\|\bm{\delta}\|_{2}+\|\bm{\beta}\|_{2}\right),
\end{align*}
where $a$ is the weight between $L_1$ and $L_2$ penalization. Cross-validation is used for selecting both tuning parameters $\lambda$ and $a$. 
 
The results of using different types of regularization are given in Table \ref{tab_l2}. We can see the clear advantage of $L_1$ penalty in error reduction, especially in longer horizon forecasts. Unlike the $L_1$ regularization, the $L_2$ penalty will not produce zeros in the fitted coefficients. That is, for those noisy search terms that are not relevant in the prediction, they will remain in the model with small, non-zero coefficients, which may lead to more variability in the prediction. In contrast, with the $L_1$ penalty, the noise and less relevant information from these search terms tend to be eliminated while only the most predictive search terms are selected to remain in the model. Similarly, although the elastic net in theory is more flexible with both $L_1$ and $L_2$ penalties, in practice it is not doing as well as the $L_1$ penalization due to the noisy search terms. The results here thus show that the noise reduction by $L_1$ penalty is quite effective in improving the performance.

\begin{table}[!h]
\footnotesize
\centering
\caption{The RMSE of PRISM under different regularization over $2007-2016$. RMSE is measured relative to the respective error of the naive method. The boldface highlights the best performance for each forecasting horizon. }\label{tab_l2}
\begin{tabular}{lrrrr}
  \hline
 & real-time & forecast 1 week & forecast 2 week & forecast 3 week \\ 
  \hline
  PRISM with $L_2$ & 0.543 & 0.551 & 0.533 & 0.543 \\ 
PRISM with Elastic Net &  0.497 & 0.490 & 0.472 & 0.483 \\ 
  PRISM & \textbf{0.493} & \textbf{0.484} & \textbf{0.461} & \textbf{0.470} \\ 
 naive & 1 (50551) & 1 (62227) & 1 (69747) & 1 (73527) \\  
   \hline
\end{tabular}
\end{table}

\section{Normality of Residuals}\label{supp_res}

\begin{figure*}[!h]
\centering
\includegraphics[width=\textwidth]{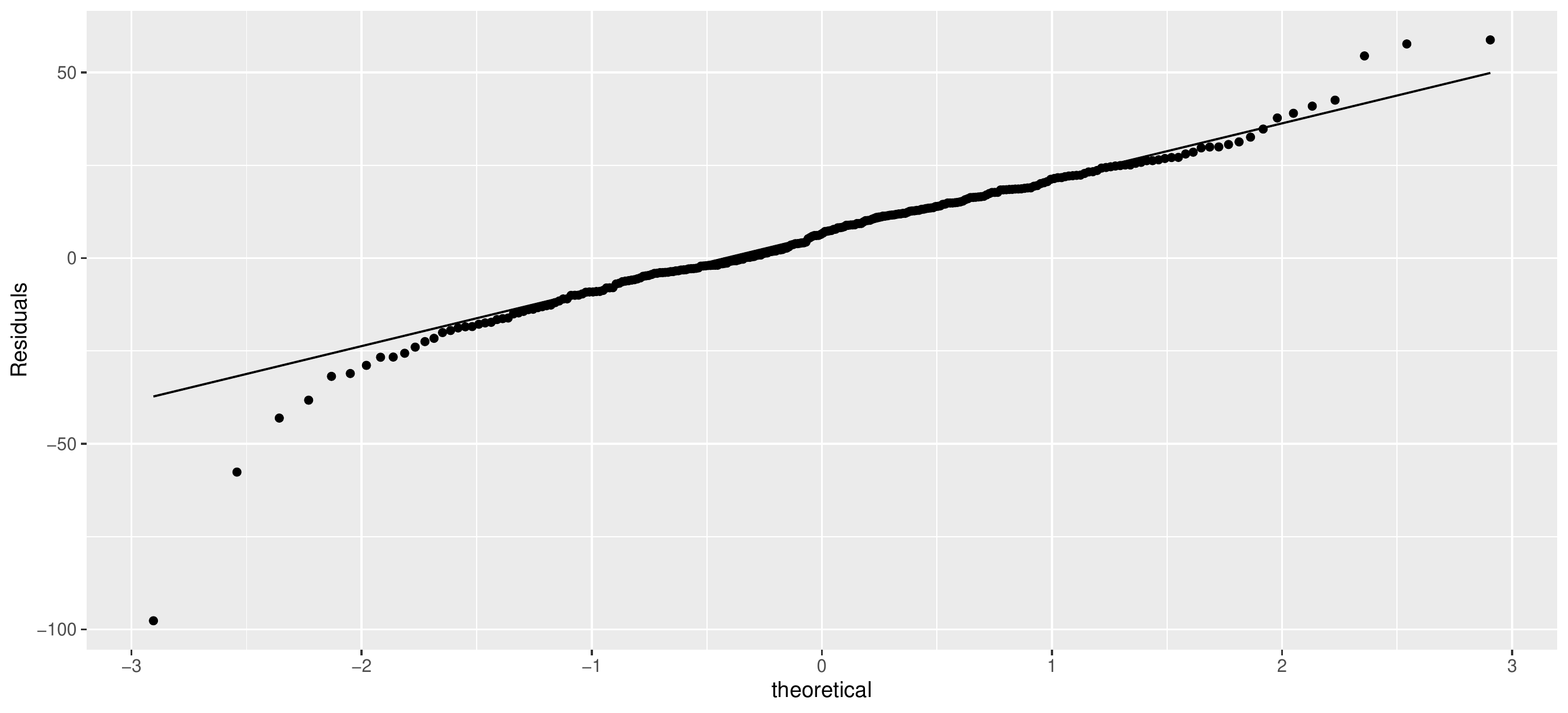}
\caption{The normal Q-Q plot of fitted residuals by PRISM in $2008-2016$.}
\label{res_qqnorm}
\end{figure*}

Here we provide empirical evidence on the normality of the residuals for constructing the predictive intervals. Figure \ref{res_qqnorm} shows the normal Q-Q plot of the fitted residuals by PRISM from 2008-2016. Except for a few points, the vast majority of the fitted residuals fall along with the normal distribution, which supports the construction of the predictive intervals by PRISM in Section 2.8 of the main text.

\begin{figure*}[!h]
\centering
\includegraphics[width=0.8\textwidth]{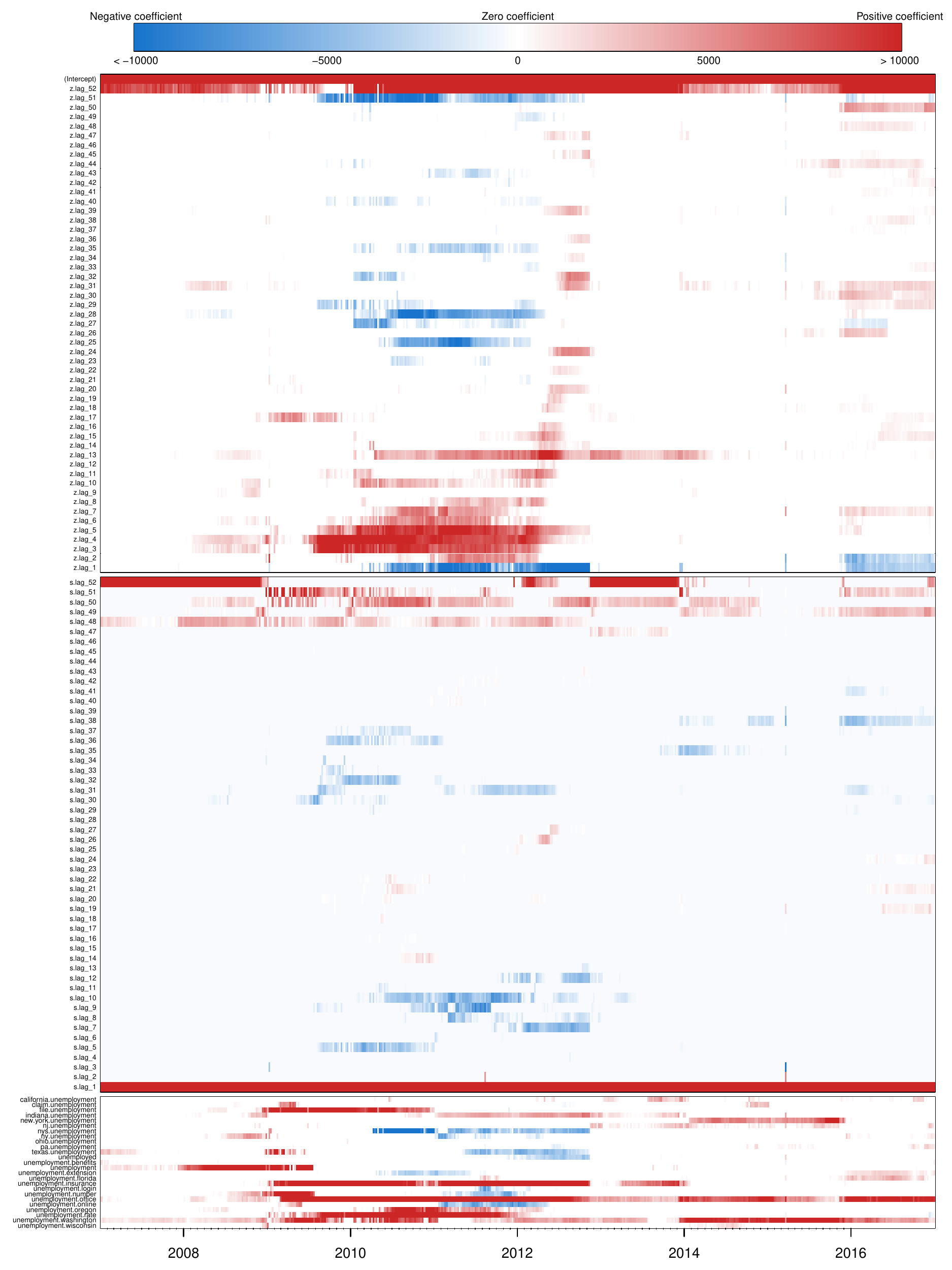}
\caption{The heatmap illustrating the dynamically fitted coefficients by PRISM for each week of forecasting in $2008-2016$. Red color represents positive coefficients, blue color represents negative coefficients, and white color represents zero.} \label{fig_coef}
\label{heatmap_coef}
\end{figure*}

\section{Fitted Coefficients of PRISM}\label{supp_coef}
We report the dynamically fitted coefficients by PRISM from each week's 
prediction in 2008-2016 in Figure \ref{fig_coef}. Focusing on the coefficients of Google search terms (the bottom section of the heatmap), we can see the sparsity of coefficients from the $L_1$ penalty. On average, $6.20$ out of the $25$ search terms are selected and included in the each week's model by PRISM during 2008-2016. Note that all 25 terms have been included at some point. In addition, we can also see different search terms come in and out of the model over time, indicating the dynamic movement for each term in its contribution to the final forecasting.

\section{Comparison with Additional Benchmarks}\label{supp_comparison}

\begin{table}[!h]
\footnotesize
\begin{center}
\begin{adjustbox}{center}
\begin{tabular}{lllllll}
\hline
 & real-time & 1 week & 2 weeks & 3 weeks \\ 
\hline
RMSE\\
\qquad  PRISM & \textbf{0.493} & \textbf{0.483} & \textbf{0.461} & \textbf{0.470} \\ 
\qquad Seasonal AR & 0.998 & 0.854 & 0.816 & 0.832 \\ 
\qquad D\text{'}Amuri and Marcucci (2017) & 1.281 & 1.089 & 1.001 & 0.971 \\
\qquad naive & 1 (50551) & 1 (62227) & 1 (69747) & 1 (73527) \\ 
MAE\\
 \qquad PRISM & \textbf{0.539} & \textbf{0.517} & \textbf{0.476} & \textbf{0.460} \\ 
\qquad Seasonal AR & 1.051 & 1.083 & 1.062 & 1.082 \\ 
 \qquad   D\text{'}Amuri and Marcucci (2017) & 1.074 & 1.022 & 0.990 & 0.915 \\ 
\qquad  naive & 1 (33637) & 1 (41121) & 1 (47902) & 1 (52794) \\  
  \hline
\normalsize
\end{tabular}
\end{adjustbox}
\caption{Performance of PRISM and additional methods over $2007-2016$ for four forecasting horizons: real-time, 1 week, 2 weeks and 3 weeks ahead. RMSE and MAE here are relative to the error of naive method; that is, the number reported is the ratio of the error of a given method to that of the naive method; the absolute RMSE and MAE of the naive method are reported in the parentheses. The boldface indicates the best performer for each forecasting horizon and each accuracy metric.}
\label{additional2016}
\end{center}
\end{table}

\begin{table}[!h]
\footnotesize
\begin{center}
\begin{adjustbox}{center}
\begin{tabular}{llllll}
 \hline
 & real-time & 1 week & 2 weeks & 3 weeks \\ 
  \hline
\qquad Seasonal AR & $1.11 \times 10^{-11}$ & $5.43 \times 10^{-13}$ & $2.82 \times 10^{-14}$ & $4.66 \times 10^{-17}$ \\ 
\qquad D\text{'}Amuri and Marcucci (2017) &  $8.58  \times 10^{-3}$& $7.64 \times 10^{-3}$ & $8.24 \times 10^{-6}$ & $4.89  \times 10^{-4}$ \\ 
   \hline
   \normalsize
\end{tabular}
\end{adjustbox}
\caption{P-values of the Diebold-Mariano test for prediction accuracy comparison between PRISM and the alternatives over 2007--2016 for four forecasting horizons: real-time, 1 week, 2 weeks and 3 weeks ahead. The null hypothesis of the test is that PRISM and the alternative method in comparison have the same prediction accuracy in RMSE.}
\label{dmtest_additional}
\end{center}
\end{table}

In this section, we compare PRISM with two additional methods: the seasonal AR model and an AR model with a single Google search term as the exogenous variable \citep{d2017predictive}. We fitted the seasonal AR with R package \texttt{forecast} \citep{hyndman2007automatic} where the order for AR component is selected by BIC. For the method of \cite{d2017predictive}, following the procedure specified in the paper, the Google search volume of ``jobs'' was included  as the exogenous variable. Table \ref{additional2016} summarizes the comparison result. It shows that PRISM leads both methods by a large margin. To assess the statistical significance of the improved prediction power of PRISM over the alternative methods, the p-values of the Diebold-Mariano test \citep{Diebold1995} are reported in Table \ref{dmtest_additional} (the null hypothesis being that PRISM and the alternative method in comparison have the same prediction accuracy in RMSE). With all the p-values smaller than 0.008, Table \ref{dmtest_additional} shows that the improved prediction accuracy of PRISM over the seasonal AR model and the method of \cite{d2017predictive} is statistically significant in all of the forecasting horizons evaluated.

\section{Comparing PRISM with Benchmarks in CSSED}\label{supp_cssed}

\begin{figure*}[!h]
\centering
\includegraphics[width=\textwidth]{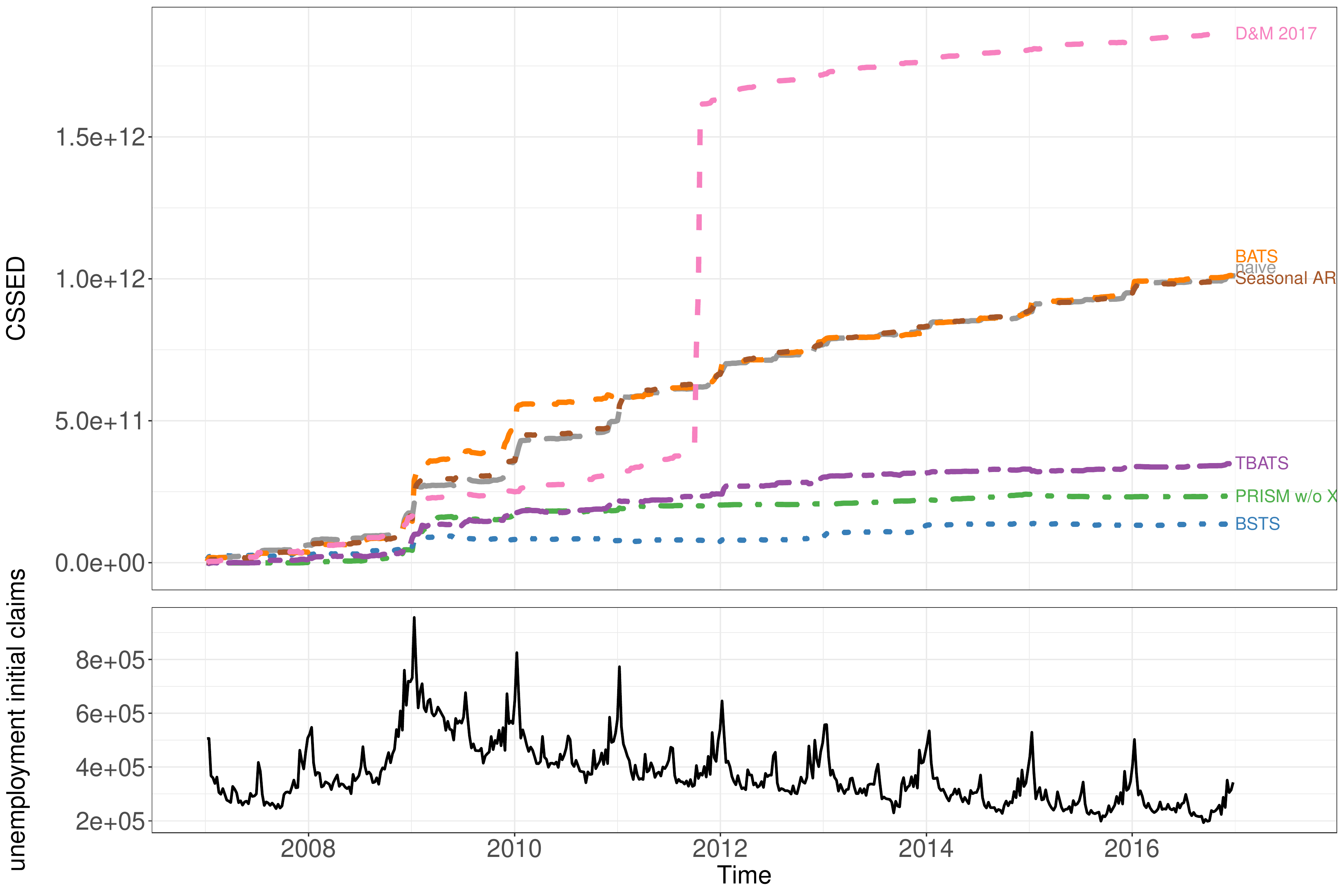}
\caption{(Top) The Cumulative
Sum of Squared forecast Error Differences (CSSED) of nowcasting between different benchmarks and PRISM. ``D\&M 2017'' denotes the method of \cite{d2017predictive}. (Bottom) The unemployment initial claims for the same period of $2007-2016$. }
\label{cssed}
\end{figure*}

\begin{figure*}[!h]
\centering
\includegraphics[width=\textwidth,page=1]{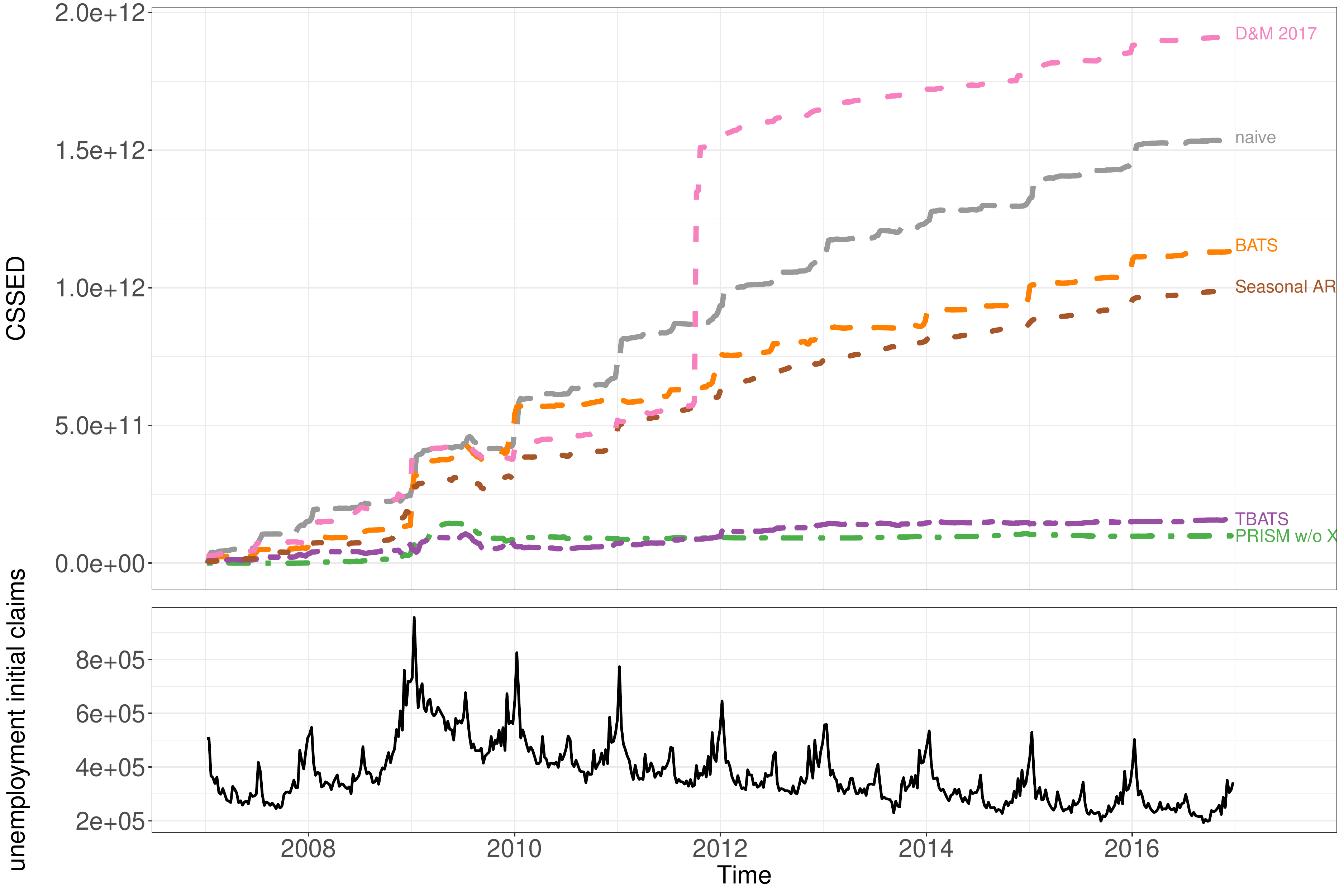}
\caption{(Top) The Cumulative
Sum of Squared forecast Error Differences (CSSED) of 1-week ahead forecasting between different benchmarks and PRISM. ``D\&M 2017'' denotes the method of \cite{d2017predictive}. (Bottom) The unemployment initial claims for the same period of $2007-2016$.}
\label{cssed_h1}
\end{figure*}

\begin{figure*}[!ht]
\centering
\includegraphics[width=\textwidth,page=3]{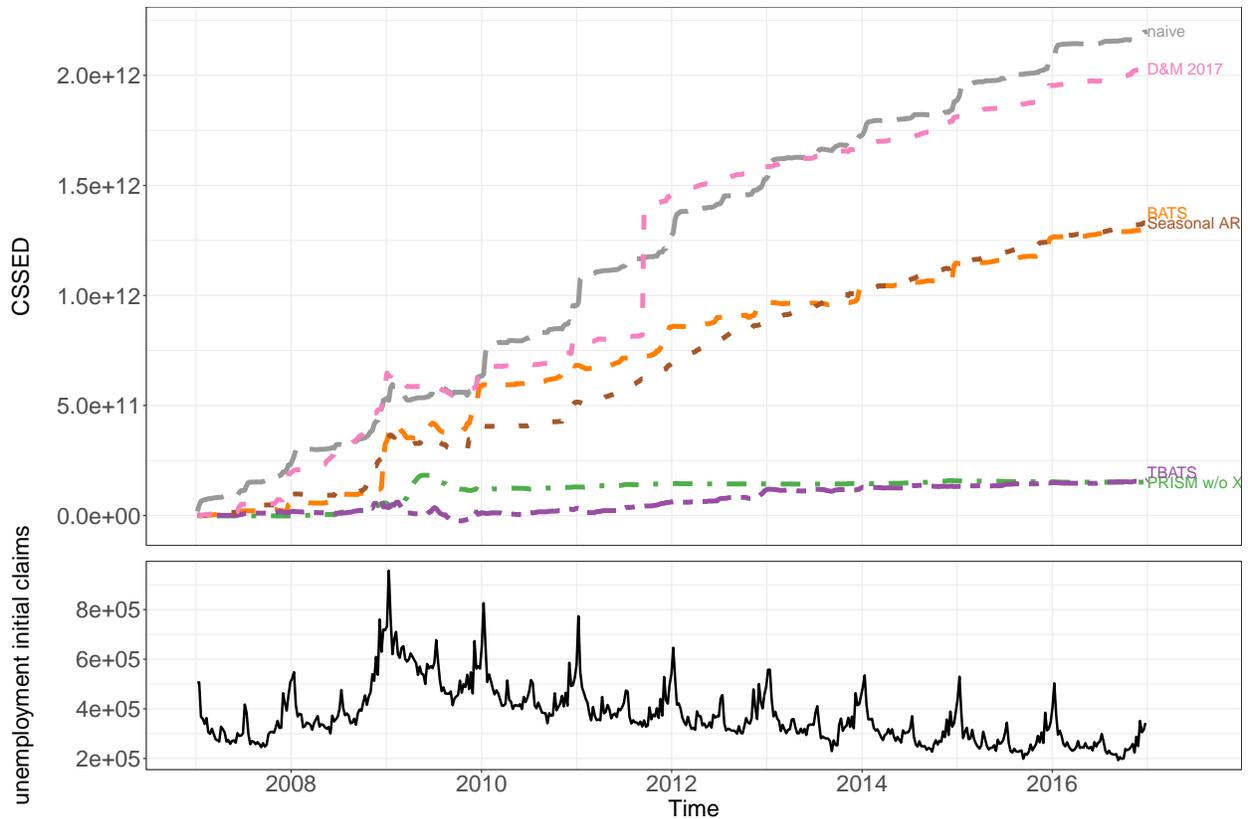}
\caption{(Top) The Cumulative
Sum of Squared forecast Error Differences (CSSED) of 3-week ahead forecasting between different benchmarks and PRISM. ``D\&M 2017'' denotes the method of \cite{d2017predictive}. (Bottom) The unemployment initial claims for the same period of $2007-2016$.}
\label{cssed_h3}
\end{figure*}

In this section, we compare PRISM with the benchmark methods according to an additional metric: Cumulative Sum of Squared forecast Error Differences (CSSED)  \cite{welch2008comprehensive}, which was recommended by a referee. The CSSED at time $T$ for benchmark method $m$ is defined by $CSSED_{m, \text{PRISM}}=\sum_{t=1}^T(e_{t,m}^2-e_{t,\text{PRISM}}^2)$, where $e_{t,m}$ and $e_{t,\text{PRISM}}$ are the prediction errors at time $t$ for the benchmark method $m$ and PRISM respectively. A positive CSSED indicates the benchmark is less accurate than PRISM. Figures \ref{cssed} - \ref{cssed_h3} showcase the CSSED between different benchmarks and PRISM over the period of 2007-2016. Consistent with the RMSE and MAE reported in Table \ref{overall} of the main text, PRISM is the leading method. 
We also see a major gain in prediction accuracy for PRISM in 2009 during the financial crisis. This also indicates the robustness of PRISM by incorporating Google search information in forecasting.

\begin{figure*}[!h]
\centering
\includegraphics[width=\textwidth,page=2]{cssed_error0719_horizons.pdf}
\caption{(Top) The Cumulative
Sum of Squared forecast Error Differences (CSSED) of 2-week ahead forecasting between different benchmarks and PRISM. ``D\&M 2017'' denotes the method of \cite{d2017predictive}. (Bottom) The unemployment initial claims for the same period of $2007-2016$.}
\label{cssed_h2}
\end{figure*}

\clearpage

\section{Year-by-year Forecasting Performance of Different Methods}\label{supp_forecast_rmse}

\begin{figure}[!h]
\centering
\includegraphics[width=\textwidth, page=1]{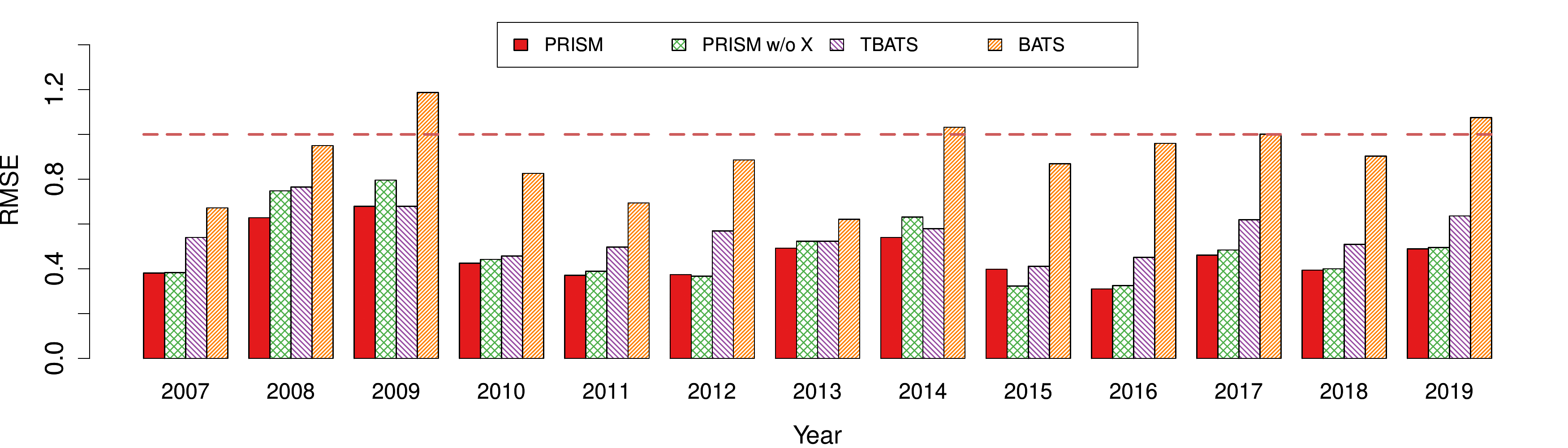}
\caption{Year-by-year 1-week ahead forecasting performance of different methods from 2007 to 2019. RMSE is measured relative to the error of the naive method; a value above 1 indicates that the method performs worse than the naive method in that time period.
}
\label{rmse_by_year_h1}
\end{figure}

\begin{figure}[!h]
\centering
\includegraphics[width=\textwidth, page=2]{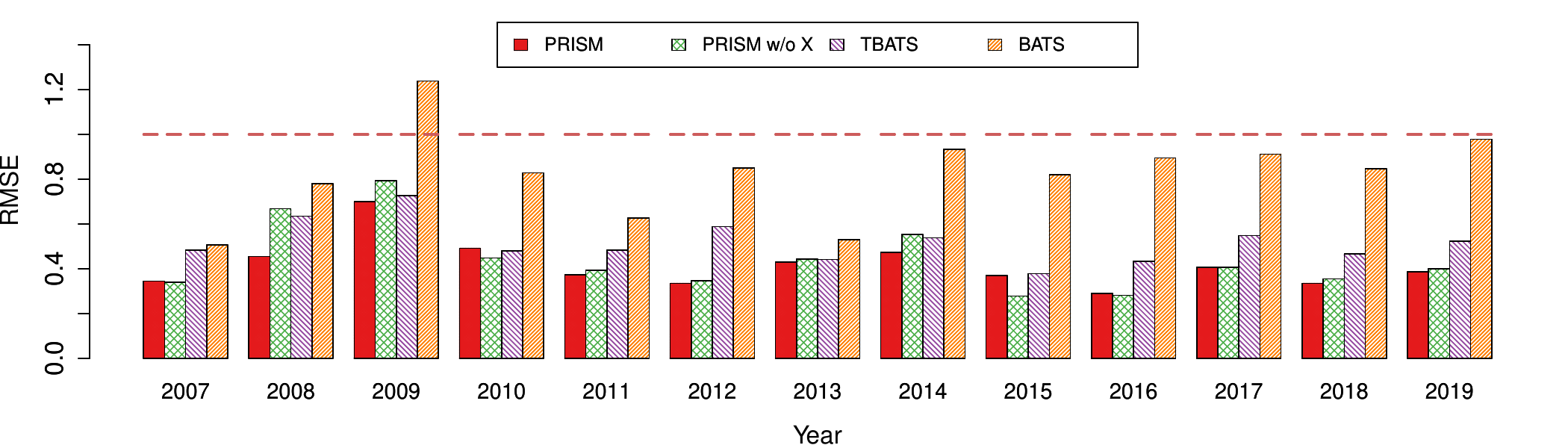}
\caption{Year-by-year 2-week ahead forecasting performance of different methods from 2007 to 2019. RMSE is measured relative to the error of the naive method; a value above 1 indicates that the method performs worse than the naive method in that time period
.}
\label{rmse_by_year_h2}
\end{figure}

\begin{figure}[!h]
\centering
\includegraphics[width=\textwidth, page=3]{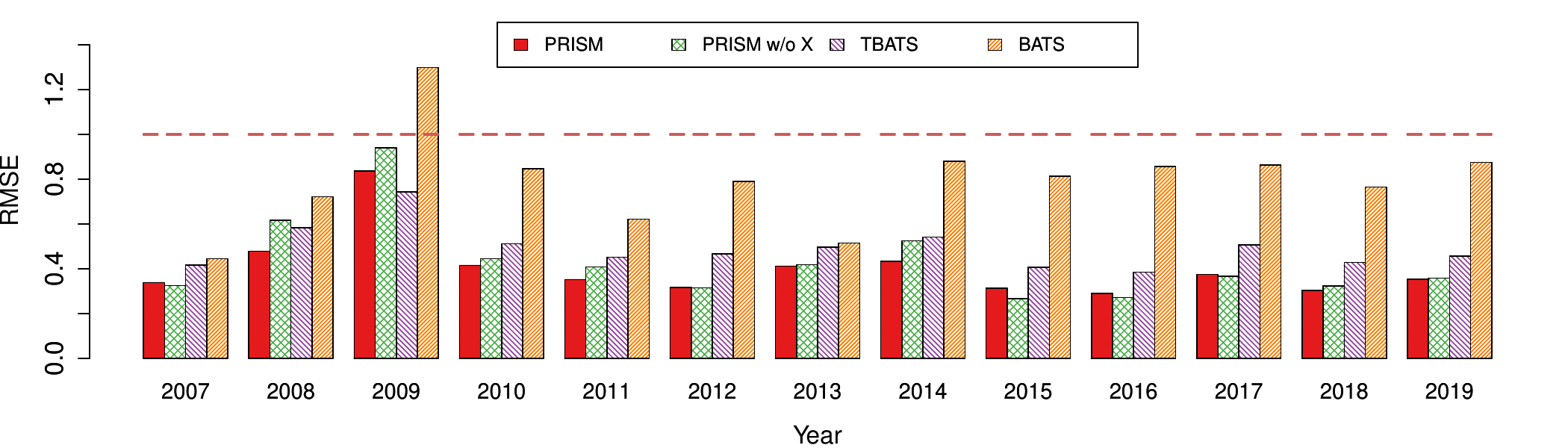}
\caption{Year-by-year 3-week ahead forecasting performance of different methods from 2007 to 2019. RMSE is measured relative to the error of the naive method; a value above 1 indicates that the method performs worse than the naive method in that time period
.}
\label{rmse_by_year_h3}
\end{figure}

In this section, we report the bar charts comparing the yearly RMSE of PRISM with other methods in longer horizon forecasting from 2007 to 2019 in Figures \ref{rmse_by_year_h1}-\ref{rmse_by_year_h3}. Note that BSTS is not in the plots as it only provides nowcasting, not forecasting (of future weeks). Similar to the nowcasting performance in Figures \ref{rmse_by_year} and \ref{rmse_by_year1719}, PRISM gives leading performance in most of the years for longer horizon forecasting.

\newpage

\section{Predictive Intervals of PRISM for Longer Horizon Forecasting}\label{supp_interval}

\begin{figure*}[!h]
\centering
\includegraphics[width=\textwidth, page=1]{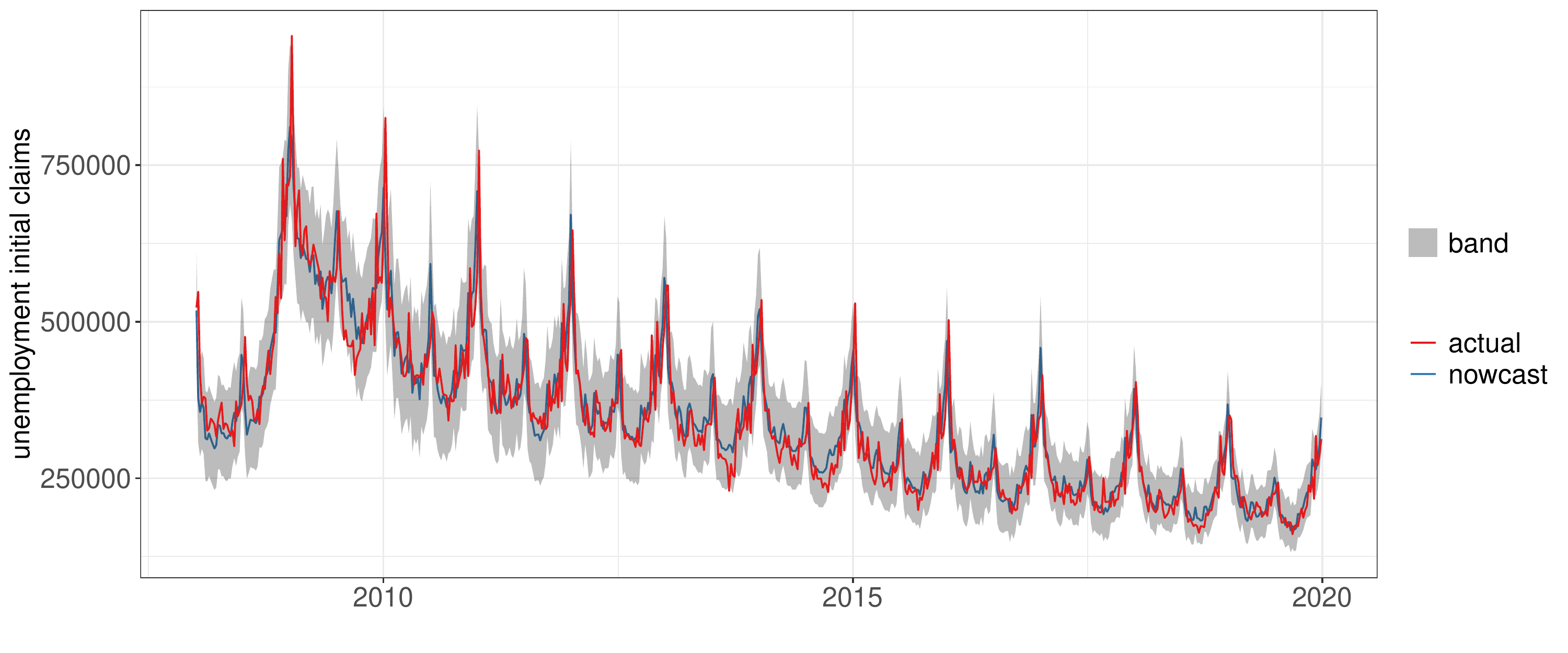}
\caption{Predictive Interval of PRISM from 2007 to 2019 for 1-week ahead forecasting. The shaded area corresponds to the 95\% point-wise predictive interval of PRISM nowcasting. The blue curve is the point estimate of PRISM nowcasting. The red curve is the true unemployment initial claims. The actual coverage of the 95\% PRISM predictive interval is 93.9\% in $2007-2019$. }
\label{sargo_err_bar_h1}
\end{figure*}

\begin{figure*}[!h]
\centering
\includegraphics[width=\textwidth, page=2]{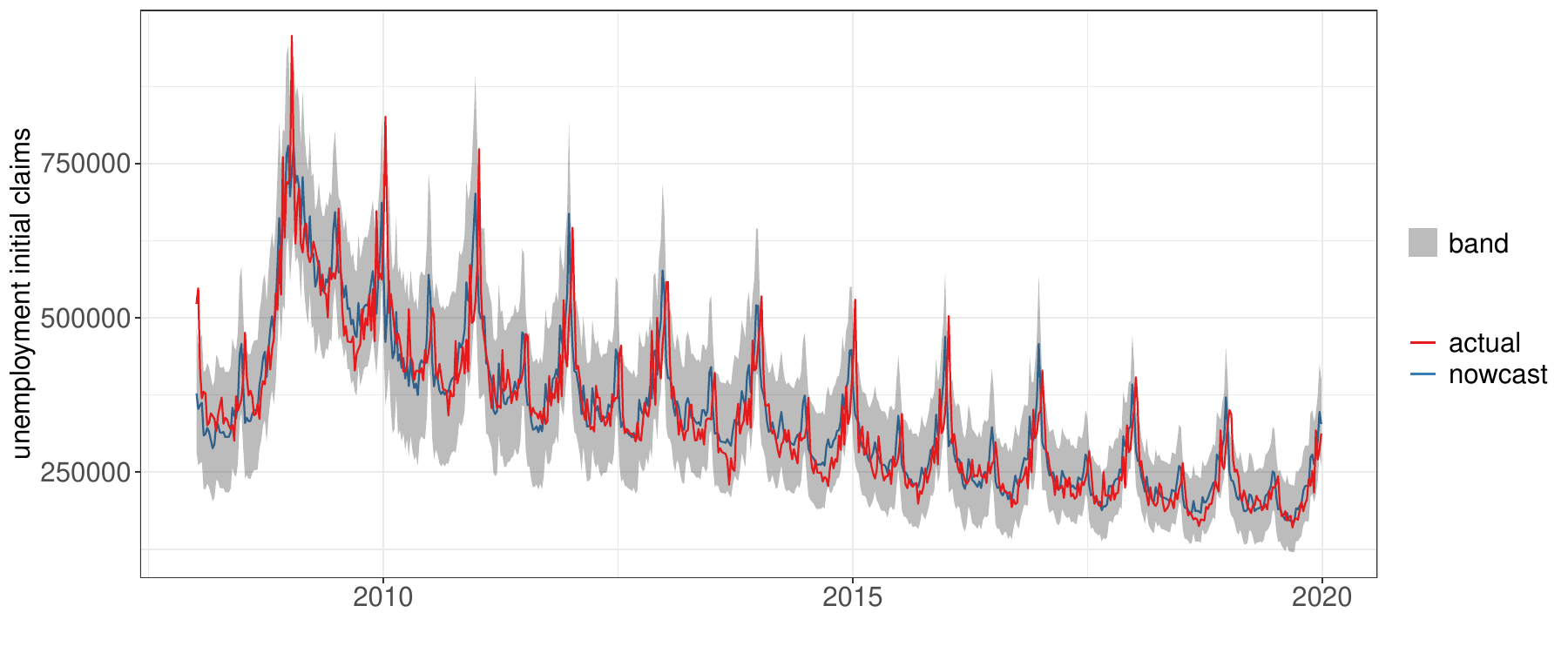}
\caption{Predictive Interval of PRISM from 2007 to 2019 for 2-week ahead forecasting.. The shaded area corresponds to the 95\% point-wise predictive interval of PRISM nowcasting. The blue curve is the point estimate of PRISM nowcasting. The red curve is the true unemployment initial claims. The actual coverage of the 95\% PRISM predictive interval is 95.4\% in $2007-2019$.}
\label{sargo_err_bar_h2}
\end{figure*}

\begin{figure*}[!h]
\centering
\includegraphics[width=\textwidth, page=3]{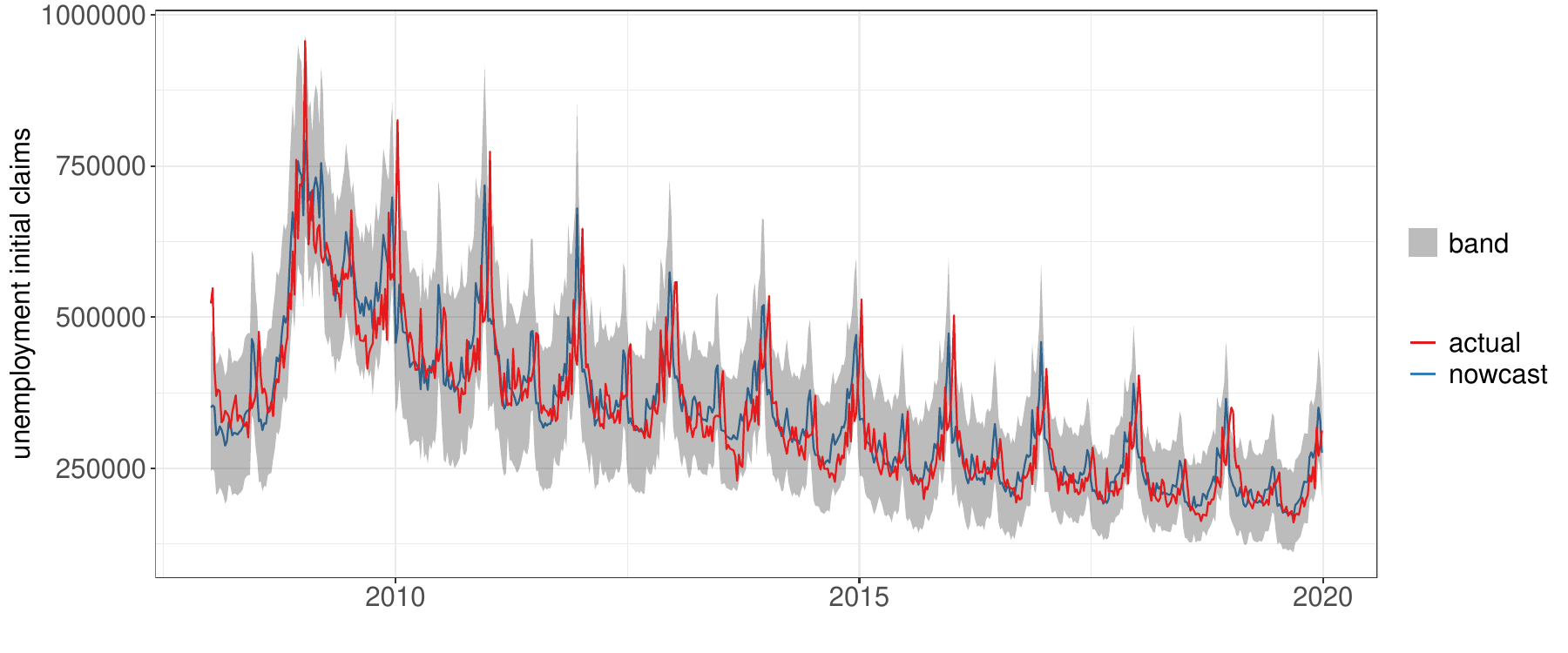}
\caption{Predictive Interval of PRISM from 2007 to 2019 for 2-week ahead forecasting.. The shaded area corresponds to the 95\% point-wise predictive interval of PRISM nowcasting. The blue curve is the point estimate of PRISM nowcasting. The red curve is the true unemployment initial claims. The actual coverage of the 95\% PRISM predictive interval is 94.7\% in $2007-2019$.}
\label{sargo_err_bar_h3}
\end{figure*}

Figures \ref{sargo_err_bar_h1}-\ref{sargo_err_bar_h3} plots the predictive intervals of PRISM in longer horizon predictions and reports the acutal coverage of the intervals. As the prediction horizon increases, the uncertainty in point prediction grows, and thereby we see the increase in the width of predictive intervals. The actual coverage of PRISM's intervals remains stable and close to the nominal 95\%: the actual coverages for 1-, 2-, 3-week ahead prediction are 93.9\%, 95.4\%, and 94.7\% respectively.

\clearpage

\section{COVID-19 Period: forecasting longer horizons}\label{supp_covid}

Table \ref{overall_covid_h} compares the performance of PRISM and the alternative methods for longer-horizon forecasting (namely, forecasting 1-week, 2-week and 3-week ahead) during the COVID-19 pandemic period. The evaluation period is from April 11, 2020 to July 18, 2020, the time period where the signal from the unprecedented jump of unemployment initial claims due to the COVID-19 pandemic trickled in and challenged longer horizon forecasting.
With input from Google search information, PRISM shows clear advantage over the alternative methods in 1-week ahead and 2-week ahead (i.e., near-future) predictions. As we are looking further into the future, the information from Google search data becomes less powerful for prediction, and the performance gap between PRISM and the other methods shrinks. This observation is consistent with what we observed in Table \ref{overall} of the main text. For 3-week ahead forecasting, PRISM is the second best, trailing TBATS (although TBATS performs quite poorly in nowcasting -- often much worse than the naive method -- as shown in Table \ref{overall_covid} and Figure \ref{rmse_covid} of the main text).

\begin{table}[!h]
\footnotesize
\begin{center}
\begin{adjustbox}{center}
\begin{tabular}{lrrrrrrr}
  \hline
& 1 week & 2 weeks & 3 weeks \\ 
  \hline
      RMSE\\
\quad  PRISM & \textbf{0.489} & \textbf{0.530} & 0.997 \\ 
\quad  BATS & 1.094 & 0.848 & 1.008 \\ 
\quad  TBATS &0.580 & 0.614 & \textbf{0.908} \\ 
\quad Seasonal AR & 1.994 & 1.522 & 1.100 \\ 
 \quad   D\text{'}Amuri and Marcucci (2017) & 5.036 & 1.792 & 1.609 \\ 
\quad  naive & 1 (873455) & 1 (1328192) & 1 (1971750)\\
   \hline
      MAE\\
\quad  PRISM & \textbf{0.427} & \textbf{0.521} & 0.917 \\ 
\quad  BATS & 0.990 & 0.733 & 0.966 \\ 
\quad  TBATS & 0.690 & 0.667 & \textbf{0.816} \\ 
\quad Seasonal AR &1.384 & 1.033 & 1.046 \\ 
 \quad   D\text{'}Amuri and Marcucci (2017) & 4.027 & 1.457 & 1.603 \\ 
\quad  naive & 1 (633992) & 1 (1003857) & 1 (1462434) \\ 
   \hline
\end{tabular}
\end{adjustbox}
\caption{Performance of PRISM and benchmark methods during COVID-19 pandemic period (April 11, 2020 to July 18, 2020) for three forecasting horizons: 1 week, 2 weeks, and 3 weeks. RMSE and MAE here are relative to the error of naive method; that is, the number reported is the ratio of the error of a given method to that of the naive method; the absolute RMSE and MAE of the naive method are reported in the parentheses. The boldface indicates the best performer for each forecasting horizon and each accuracy metric.  }\label{overall_covid_h}
\end{center}
\end{table}

\bibpunct{(}{)}{;}{a}{}{,} 
\bibliographystyle{jasa}
\bibliography{reference}

\end{document}